\documentclass[12pt]{article}
\usepackage{amssymb,amsmath,amscd}
\makeatletter

\@addtoreset{equation}{section}
\def\section{\@startsection {section}{1}{\z@}{-2.25ex plus -1ex minus
 -.2ex}{1.0ex plus .2ex}{\large\bf}}
\def\subsection{\@startsection{subsection}{2}{\z@}{-2.0ex plus%
 -1ex minus -.2ex}{0.5ex plus .2ex}{\bf}}

\def\Ad{\mathrm{Ad}}
\def\ad{\mathrm{ad}}
\def\rbr{\right)}
\def\lbr{\left(}

\def\bmu{{\mbox{\boldmath $\mu$}}}

\newcommand{\inv}[0]{{-1}}
\newcommand{\ntg}[0]{{n+2g}}

\newcommand{\cd}[0]{\!\cdot\!}

\newcommand{\flh}[3] {f_{{#1}{#2}}^{\;\;\;\;{#3}}\;}

\newcommand{\surfb}[0]{S_{g,n}^\infty}
\newcommand{\xb}[0]{x_\infty}
\newcommand{\cb}{l_\infty}
\newcommand{\rb}{P_\infty}

\newcommand{\Omb}{\Omega_\infty}
\newcommand{\pp}{p}

\newcommand{\surf}[0]{S_{g,n}}

\newcommand{\prgr}{G\ltimes \mathfrak{g}^*}
\newcommand{\pralg}{\mathfrak{g}\,\tilde \oplus\,\mathfrak{g}^*}

\newcommand{\csg}[0] {C}
\newcommand{\csalg}[0] {\mathfrak{c}}

\def\ba{{\mbox{\boldmath $a$}}}
\def\bx{{\mbox{\boldmath $x$}}}
\def\bs{{\mbox{\boldmath $s$}}}
\def\by{{\mbox{\boldmath $y$}}}
\def\bj{{\mbox{\boldmath $j$}}}

\def\pbj{{\mbox{\bf \j} }}
\def\bk{{\mbox{\boldmath $k$}}}
\def\bp{{\mbox{\boldmath $p$}}}

\newcommand{\gothg}{\mathfrak g }
\newcommand{\gothc}{\mathfrak c }

\newcommand{\RR}{\mathbb{R}}
\newcommand{\CC}{\mathbb{C}}

\newtheorem{theorem}{Theorem}[section]
\newtheorem{lemma}[theorem]{Lemma}
\newtheorem{corollary}[theorem]{Corollary}

\def\bea{\begin{eqnarray}}
\def\eea{\end{eqnarray}}
\def\bmz{\left(\begin{array}{2,2}}
\def\emz{\end{array}\right)}
\def\bmd{\left(\begin{array}{3,3}}
\def\emd{\end{array}\right)}

\newcommand{\mi}[0]{{M_i}}
\newcommand{\ai}[0]{{A_i}}
\newcommand{\bi}[0]{{B_i}}

\newcommand{\aj}[0]{{A_j}}
\newcommand{\bjj}[0]{{B_j}}

\newcommand{\me}[0]{{M_1}}

\newcommand{\mf}[0]{{M_n}}

\begin{document}
\parskip 5pt
\parindent 0pt
\begin{flushright}
EMPG-05-07\\
\end{flushright}

\begin{center}
\baselineskip 24 pt {\Large \bf Phase space structure of Chern-Simons
  theory with a non-standard puncture}

\baselineskip 18 pt

\vspace{1cm} {{ C.~Meusburger}\footnote{\tt  cmeusburger@perimeterinstitute.ca}\\
Perimeter Institute for Theoretical Physics\\
31 Caroline Street North,
Waterloo, Ontario N2L 2Y5, Canada\\

\vspace{1cm}
{ B.~J.~Schroers}\footnote{\tt bernd@ma.hw.ac.uk} \\
Department of Mathematics, Heriot-Watt University \\
Edinburgh EH14 4AS, United Kingdom } \\

\vspace{0.5cm}

{May   2005}

\end{center}

\begin{abstract}
\noindent We explicitly determine
 the symplectic structure on the phase space of
 Chern-Simons theory with gauge group $\prgr$ on a three-manifold
of topology $\RR \times \surfb$, where $\surfb$ is a surface of genus
$g$ with $n+1$ punctures. At each puncture additional variables
are introduced and coupled minimally to the Chern-Simons gauge field.
 The first $n$ punctures are treated in the usual way and
the additional variables   lie on coadjoint
orbits of $\prgr$. The $(n+1)$st puncture  plays a distinguished role
and the associated variables lie in  the cotangent bundle
of $\prgr$. This allows us to impose a curvature singularity for
the Chern-Simons gauge field at the
distinguished puncture with an arbitrary Lie algebra valued
coefficient.
The treatment of the distinguished
puncture is motivated by the desire to construct a simple model for
an open universe
 in the Chern-Simons formulation of $(2+1)$-dimensional
gravity.

\baselineskip 12pt \noindent
\end{abstract}

\section{Introduction}

Chern-Simons field theory has attracted the attention of both
mathematicians and physicists. Its relevance in mathematics is
largely due to its applications in fields  such as knot theory,
see \cite{Witten2} or the book \cite{Guadagnini}, and  the theory
of moduli spaces of flat
 connections, see \cite{Atiyah} for a summary. These research areas in turn provide useful concepts and methods for the study of Chern-Simons theory.
From a physicist's point of view, Chern-Simons theory is
interesting because it captures important aspects of real physical
systems and is at the same time mathematically tractable. While
not  describing any real physical system, it plays an important
role in, for example, the modelling of certain condensed matter
systems. More importantly,  it shares structural features with
fundamental  physical theories. Its diffeomorphism invariance, for
 instance,  makes it a useful toy model for the investigation of
 coordinate independent approaches to quantisation. In particular, it is relevant to the study of Einstein's theory
 of gravity, which in (2+1) dimensions can be formulated as
a Chern-Simons theory \cite{AT,Witten1}.

The  basic reason for the relative mathematical simplicity of
Chern-Simons theory
 is that, with appropriate boundary conditions and after dividing out
gauge degrees of freedom, it  has  a finite-dimensional phase
space. Thus, in suitable circumstances
 Chern-Simons theory allows one to reduce
the field-theoretic description of a
physical system to a mathematically well-defined  and
finite-dimensional
  model. This reduction from an infinite to a  finite
number of degrees of freedom makes the classical theory tractable
and considerably simplifies the quantisation.

In this paper we study Chern-Simons theory in its Hamiltonian formulation
on  a manifold of topology $\RR \times \surfb$,
where $\surfb$ is a surface of genus
$g$ with $n+1$ punctures. The purpose of the paper is to
introduce a new way of  treating  punctures in
Chern-Simons theory and to determine  the physical phase space and
its symplectic
structure when one distinguished
puncture is treated in this way. The main result is an explicit
determination  of  the symplectic structure on
the finite-dimensional physical phase space.

The motivation for our
treatment of the distinguished puncture comes from the
 Chern-Simons formulation of (2+1)-dimensional gravity.
As we explain in  detail in a separate paper
\cite{we5}, the distinguished puncture can be used to model
``spatial infinity'' in open universes.
 Applied to (2+1)-dimensional gravity, our model leads
to  a finite-dimensional description of the phase space which can
serve as a starting point for both an investigation of the
classical dynamics and for quantisation, using the methods
developed in \cite{we2}.
 The relation to (2+1)-dimensional gravity is also the
motivation  for our choice of gauge group. We consider gauge groups
of the form $\prgr$, where $G$ is an arbitrary finite-dimensional
Lie group.  They include as special cases the Euclidean group
 and the Poincar\'e group in three dimensions which arise in the
Chern-Simons formulation of
 Euclidean and Lorentzian (2+1)-gravity with vanishing cosmological
 constant
\cite{Witten1}. Mathematically, groups of the form $\prgr$ are
particularly simple examples of Poisson-Lie groups, which is
important in our analysis. We stress, however, that our treatment
of the distinguished puncture and many of our results concerning
the phase space are not limited to gauge groups of this type.

Since much of the paper is quite technical, we give a brief sketch
of our treatment of punctures on $\surfb$. The usual  approach,
followed in \cite{Witten2} and summarised in \cite{Atiyah}, is to
require  the curvature of the gauge field to have a delta-function
singularity on the line $\RR \times \{x_{(i)}\}$, where $x_{(i)}$
is the coordinate of the puncture on $\surfb$, and to restrict the
Lie algebra valued coefficient of the delta-function to a fixed
coadjoint orbit of the gauge group. In order to achieve this,
additional variables need to be introduced which parametrise the
coadjoint orbit. The dynamics of these additional variables is
governed by the Kirillov-Kostant-Souriau symplectic structure on
the coadjoint orbit  minimally coupled to the Chern-Simons gauge
field. In this paper we treat the first $n$ punctures on $\surfb$
in this way. At the  distinguished puncture, whose coordinate on
$\surfb$ is denoted $x_\infty$, we also require the curvature to
have a delta-function singularity, but this time we do not
restrict the Lie algebra valued coefficient $T$. Instead we
introduce an additional Lie group valued variable $g$ at the
distinguished puncture and interpret the pair $(g,T)$ as an
element of the cotangent bundle of the gauge group. The
dynamics of the  variables $g$ and $T$ is governed by the canonical
symplectic structure on the cotangent bundle minimally coupled to
the Chern-Simons gauge field.

Having defined the field-theoretical model we parametrise the
physical phase space as a finite-dimensional quotient of a
finite-dimensional extended phase space and  compute the pull-back
of the
 symplectic structure to the extended phase space.
Our calculations  make extensive use of the results and techniques
of \cite{AMII}. In that paper, Alekseev and Malkin consider
Chern-Simons theory with a simple, complex  gauge group (or its
compact  real form)  on a three-manifold of the form  $\RR \times
\surf$, where $\surf$ is a surface of genus $g$ with  $n$
punctures which are treated in the usual way. Here we extend their
analysis to include the distinguished puncture and to apply  to
gauge groups of the form $\prgr$, where $G$ is any real Lie group.
In particular, we do not assume the  existence of  a single
abelian Cartan subalgebra into which any  Lie algebra  element can
be conjugated; such a Cartan subalgebra plays an important role in
\cite{AMII}. However, the semi-direct product structure of $\prgr$
also leads to simplifications. It  allows us to give a more
detailed  formula for pull-back of the
 symplectic form to the extended phase space and to show that this pull-back is exact.

The paper is organised as follows. After explaining our notation
and conventions in  Sect.~2, we define our field-theoretical model
by giving the action and equations of motion in Sect.~3. The
rather technical Sect.~4 explains the adaptation of  Alekseev and
Malkin's method to our situation and ends in a first formula for
the pull-back of the symplectic form to the extended phase space,
valid for any gauge group. In Sect.~5 we use  the structure of the
gauge group $\prgr$ to derive the formula for the pull-back of the
symplectic form as an exterior derivative of a one-form. In
Sect.~6 we explain in detail how  the physical phase space is
obtained from the extended phase space by imposing various
constraints and dividing by the associated  gauge transformations.
Sect.~7  contains  a short discussion and conclusion, and the
Appendices \ref{heisenberg} and \ref{alekseevmalkin},
respectively, summarise the properties of groups $\prgr$ as
  Poisson-Lie groups and key results from
\cite{AMII} needed in the main part of the paper.


\section{Setting and Notation}

Let $M$ be a  three-manifold  of topology
 $\RR\times\surfb$, where $\surfb$ is an oriented two-dimensional
manifold  of genus $g$ with $n+1$ punctures.
Of these,  one will play  a special role, and we refer to it as the
distinguished puncture to differentiate it from the remaining $n$
ordinary punctures.
We introduce a global coordinate $x^0$ on $\RR$,
write  $x=(x^1,x^2)$ for local coordinates on the surface
$\surfb$ and denote the differentiation with respect to $x^0, x^1$ and
$x^2$ by $\partial_0,\partial_1$ and $\partial_2$.
The  coordinates of the $n$ ordinary punctures on $\surfb$
are  $x_{(1)},\ldots,x_{(n)}$, and the
distinguished puncture has the
coordinate $\xb$. We refer to  the one-dimensional
submanifolds defined by $x=x_{(i)}$, $i=1,\ldots, n$,  and $x=\xb$ as,
respectively, the  world lines of
the ordinary punctures and the distinguished  puncture.

Chern-Simons theory is a field theory for an $H$-connection on
$M$. For now, $H$ can be an arbitrary finite-dimensional Lie
group, but we will restrict ourselves to a particular class of Lie
groups further below. Locally, the connection is given by a Lie algebra valued  one-form
$A$ on $M$, called the gauge field.
The product structure  $M=\RR\times\surfb$ allows us to
decompose the gauge field as
\begin{align}
\label{gauge field}
A=A_0 dx^0+A_S,
\end{align}
where $A_S$ is an $x^0$-dependent and Lie algebra valued
one-form on $\surfb$ and $A_0$ is a Lie algebra valued function on
$\RR\times\surfb$.
In the following we
denote by $d$ the  exterior derivative  on $\RR\times\surfb$. Occasionally, we need to differentiate with respect to the dependence on $\surfb$ only
and denote such derivatives by $d_S$.
With this notation, the curvature  of the connection $A$ is the two-form
\bea
\label{curv}
F=dA + A\wedge A,
\eea
and the splitting \eqref{gauge field} leads to the decomposition
\bea
\label{curvdec}
F=dx^0\wedge(\partial_0 A_S - d_S A_0 + [A_0,A_S]) + F_S,
\eea
where $F_S$ is the  curvature two-form on $\surfb$:
\bea
\label{curvv}
F_S=d_SA_S + A_S\wedge A_S.
\eea

In the following, we consider gauge groups that are
semi-direct
products
 $H=\prgr$   of a connected,
finite-dimensional
 Lie group $G$ and the dual $\gothg^*$ of its  Lie algebra
$\gothg=\text{Lie}\,G$, viewed as an abelian group.
The group $G$ acts on its Lie algebra $\gothg$ by the adjoint action $\Ad$
and,
following the conventions of \cite{MaRa},
 we define $\Ad^*(g)$ as the algebraic dual of $\Ad(g)$, i.~e.~
\bea
\label{coadact} \langle \Ad^*(g)\bj, \xi\rangle=\langle \bj, \Ad(g)\xi \rangle\qquad\forall \bj\in\gothg^*, \xi\in\gothg, g\in G,
\eea
where  $\langle,\rangle$ is the canonical pairing of elements of $\gothg$ with elements of $\gothg^*$.
Note that with this definition
the coadjoint action of $g\in G$ is given by $\Ad^*(g^\inv)$.
Writing elements of $ \prgr$ as $(u,\ba)$ with $u\in G$, $\ba\in\gothg^*$,
 the group multiplication in $\prgr$  is
\bea
\label{groupmult} (u_1,\ba_1)\cdot(u_2,\ba_2)=(u_1\cdot u_2,\ba_1+\Ad^*(u_1^\inv)\ba_2).
\eea

In this paper, all Lie algebras will be considered over $\RR$.  The
Lie algebra of $\prgr$ is $\gothg\oplus\gothg^*$ as a vector space,
and we denote it as $\pralg$ to remind the reader that it is not the
direct sum of $\gothg$ and $\gothg^*$ as a Lie algebra.
The Lie bracket can be characterised
as follows. Let
$J_a$, $a=1,\ldots,\text{dim}\,G$, be
 a basis  of the Lie algebra $\gothg$ and let  $P^a$, $a=1,\ldots,
\text{dim}\,G$,
be the dual basis of $\gothg^*$ i.e. $\langle P^b, J_a \rangle=\delta_a^b$.
Then  $J_a, P^b $,
$a,b=1,\ldots,\text{dim}\,G$, form a basis of $\pralg$
with commutators
\bea
\label{commutator} [J_a,J_b]=\flh a b c
J_c\qquad[J_a,P^b]=-\flh a c b P^ c\qquad[P^a,P^b]=0,
\eea
where $\flh a b c$ are the structure constants of $\gothg$.

The pairing  $\langle,\rangle$ between $\gothg$ and $\gothg^*$
can be extended to a symmetric,
non-degenerate bilinear form on  $\pralg$,
also denoted  by $\langle,\rangle$, by setting
\bea
\langle \xi,\bj \rangle = \langle \bj, \xi \rangle,\quad \mbox{and}\quad
 \langle \bj,\bk \rangle =\langle \xi, \eta \rangle =0, \quad \bj,\bk
\in \gothg^*, \,\,\xi,\eta \in \gothg. \eea This pairing  is
$\prgr$-invariant by virtue of \eqref{coadact}, and with this
pairing the vector space $\gothg\oplus\gothg^*$ is canonically
isomorphic to its dual. We will  use this isomorphism
 to identify $\pralg$ with  its dual  without writing it explicitly
 in the following; both will be denoted by  $\pralg$.
In particular,
 we write  both the adjoint and the coadjoint
action of an element $h\in\prgr$  on
$\pralg $  simply by conjugation with $h$.

In our discussion of the behaviour of  the Chern-Simons gauge
field at the puncture we make use of coadjoint  or, equivalently, adjoint
orbits of the $\prgr$. It is then convenient to parametrise these
orbits in the form $\{hDh^\inv|h\in\prgr\}$, where $D$ is a fixed
element of the Lie algebra. When dealing with Lie algebras
over $\CC$ or compact forms of  complex Lie algebras as in \cite{AMII},
 it is
possible to fix one Cartan subalgebra of the Lie algebra and choose
$D$ to lie in that Cartan subalgebra without loss of
generality. However, for  general Lie algebras over $\RR$
the theory of Cartan subalgebras, defined as nilpotent subalgebras which
are their own normaliser, is more complicated. There are two features which
are important for us.

The first is  related to the conjugacy of
Cartan subalgebras.
As explained in \cite{Varadarajan} or \cite{Knapp},  Cartan subalgebras of
real Lie algebras  can not necessarily be conjugated into
 each other. Instead,
there exists a family of Cartan subalgebras
$\gothc_\iota$,  $\iota\in I $,  where $I$ is
some finite index set, such that any Cartan subalgebra
is conjugate to one of the $\gothc_\iota$. Moreover, not all elements of
the Lie algebra lie in a Cartan subalgebra. This is only the case for
regular elements\footnote{By definition, a regular element $T$ of a Lie
  algebra  is
such that the multiplicity of the characteristic root of $\ad T$
is equal to the rank of the Lie algebra.}. A regular element can
therefore always be written as $T=hDh^\inv$, where $D$ is in one
of  a finite number
  of fixed Cartan subalgebras.

The second issue we need to be aware of is that  a Cartan
subalgebra of a real Lie algebra  is not necessarily abelian.
In order to make sure that all Cartan subalgebras of $\prgr$ are
abelian  we need to assume that $G$ is semi-simple. This can be
seen as follows. Cartan subalgebras of semi-simple Lie algebras
are abelian \cite{OV}. Furthermore, if we assume that $\gothg
$ is semi-simple it follows from Theorem 9.5 in  chapter 1 of \cite{OV} applied
to the Levi decomposition $\pralg$
that Cartan subalgebras of $\pralg$
are of the form $\gothc\oplus\gothc^*$, with $\gothc$ being a
Cartan subalgebra of $\gothg$ and $\gothc^*$ a Cartan subalgebra
of $\gothg^*$. However,  such subalgebras of $\pralg$ are
automatically abelian.

With these structural features in mind,
we adopt the following conventions for Cartan subalgebras.
In order to parametrise regular elements
we pick a finite family of  Cartan subalgebras $\gothc_\iota, \iota
\in I$ of
$\pralg$ for  and
write $T_{\gothc_\iota}$ for the  subgroup
obtained by exponentiating the Cartan subalgebra
$\gothc_\iota$. We assume the existence of at least one
abelian Cartan subalgebra and denote it by $\gothc$; the
abelian subgroup obtained by exponentiating $\gothc$ is denoted
$T_\gothc$.

The group $\prgr$  has the structure of a Poisson-Lie group, which is
described in detail in \cite{we2} and partly reviewed in  Appendix
\ref{heisenberg}.
Poisson-Lie groups have compatible Poisson and Lie group structures,
and for every Poisson-Lie group there is a dual Poisson-Lie group
where Lie  and Poisson structure are, in a suitable sense,
interchanged.
 As a group, the dual of $\prgr$ is the direct product $G\times \gothg^*$. The
Poisson-Lie group structure gives rise to a diffeomorphism between
$\prgr$ and its dual $ G\times \gothg^*$ which is given by
\eqref{diffeo} in Appendix \ref{heisenberg} . The practical use of
this diffeomorphism for our calculations is a parametrisation of
elements in $\prgr$ in terms of  elements of $G\times \gothg^*$.
Writing an element of $G\times \gothg^*$
 as $(u, -\bj)$ the corresponding element in $\prgr$ is given by
\bea
\label{gparam} (u,\ba)=(u,-\Ad^*(u^\inv)\bj)\qquad\text{with}\; u\in
G,\;\ba,\bj\in\gothg^*.
\eea

\section{Action and  equations  of motion }

In order to define Chern-Simons theory with gauge group $\prgr$ on
a surface with punctures we  need  to specify the behaviour of the
gauge field near the punctures. The standard way of including
punctures in Chern-Simons theory
  is to allow the gauge field to have singularities of a particular
form at the punctures.
The singularities are such that the curvature \eqref{curvv} of the spatial gauge field $A_S$
develops delta-function singularities
with Lie algebra valued coefficients restricted to fixed  coadjoint
orbits of $\prgr$. However, motivated by the application to the
Chern-Simons formulation of (2+1)-dimensional gravity, see \cite{we5},
we now formulate an action functional which allows for the
inclusion of an additional puncture where the
 restriction to fixed coadjoint orbits is not imposed.

We start by summarising  the usual treatment of punctures in
Chern-Simons theory \cite{Witten2},
adapted to gauge groups of the form $\prgr$.
 For Chern-Simons theory on a genus $g$ surface $S_{g,n}$ with $n$ punctures at coordinates  $x_{(1)},\ldots,x_{(n)}$,
 one requires that the spatial curvature \eqref{curvv} takes the form
\begin{align}
\label{vara00equ} &\tfrac{k}{2\pi}F_S(x)=\sum_{i=1}^n
T_i\delta^{(2)}(x-x_{(i)})dx^1\wedge dx^2
\end{align}
with Lie algebra valued coefficients $T_i\in \pralg$ restricted to
 certain coadjoint orbits of $\prgr$.
This restriction is imposed to ensure that the phase space of the theory is symplectic
\cite{Witten2}. We assume in the following  that the  elements $T_i$
are regular. This assumption simplifies the notation and saves us
having to  distinguish cases in the discussion. Moreover, it is
satisfied in the application to (2+1)-dimensional gravity
which is our main motivation.
 As explained in the
previous section, each regular element $T_i$
  can then be conjugated into one of a  family
of Cartan subalgebras $\gothc_\iota$, with $\iota$ ranging over a finite index
set $I$.  Thus we write each $T_i$
in terms of  $h_i\in \prgr$ and a fixed elements
 $D_i\in \cup_{\iota \in I} \gothc_\iota$ as
\bea
\label{tidef} T_i=h_i D_i h_i^\inv.
\eea
 For a more detailed
description and an explicit parametrisation of coadjoint orbits of
$\prgr$ we refer the reader to Appendix \ref{cotbundle}.

The action for Chern-Simons theory coupled to punctures
contains  the Chern-Simons action for the gauge field $A$
 and  kinetic terms for
the orbit  variables $h_i\in\prgr$ which are derived from the
Kirillov-Kostant-Souriau symplectic structure on the coadjoint
orbits and coupled to the gauge field via minimal coupling. The
product structure $\RR\times \surf$  allows us to give the action
in its Hamiltonian form, which makes use of the decomposition
\eqref{gauge field}:
\begin{align}
 S[A_S,A_0, h_i]=
\label{ordaction}
 &\int_\RR dx^0\int_{\surfb}\tfrac{k}{4\pi}\langle\partial_0 A_S\wedge A_S\rangle
 -\int_\RR dx^0 \sum_{i=1}^n \langle D_i\,,\, h_i^\inv\partial_0 h_i\rangle\\
 +&\int_\RR dx^0\int_{\surfb}\langle A_0\,,\, \tfrac{k}{2\pi}F_S -
\sum_{i=1}^n T_i\delta^{(2)}(x-x_{(i)})dx^1\wedge dx^2 \rangle.
\nonumber
\end{align}
Note that $A_0$ plays the role of a Lagrange multiplier and that
the fixed elements  $D_i$ enter as parameters.

We now include an additional distinguished puncture
 labelled by $\infty$, at which
the curvature also develops a Lie algebra valued singularity, but
this time without restriction on the Lie algebra element
multiplying the delta-function. In order to impose the desired
curvature singularity we introduce a further $\pralg$-valued
variable $T$ (not restricted to lie on a particular coadjoint
orbit) as well as a group valued variable $g\in \prgr$ which will
play the role of the conjugate variable to $T$. Similar to the
case of the ordinary punctures, the component $A_0$ should again
act as a Lagrange multiplier imposing the constraint on the
spatial curvature \eqref{curvv}, and there should be a dynamical
term involving the variables $T$  and $g$. However, in contrast to
the ordinary punctures, where the kinetic terms are obtained from
the canonical symplectic potential for coadjoint orbit of $\prgr$,
the kinetic term we propose for the distinguished puncture is
obtained from the canonical symplectic potential on the cotangent
bundle of $\prgr$. Both these symplectic potentials are discussed
and derived in the Appendix  \ref{cotbundle}.

The full action for  Chern-Simons theory on
$\RR\times \surfb$ then depends on the gauge field $A$, the
group valued functions $h_i,g$ of $x^0$ and the Lie algebra valued
function $T$ of $x^0$ and is given by
\begin{align}
 &S[A_S,A_0, h_i, T,g]=
\label{action}\\
 &\quad\;\;\int_\RR dx^0\int_{\surfb}\tfrac{k}{4\pi}\langle\partial_0 A_S\wedge A_S\rangle
 -\int_\RR dx^0 \sum_{i=1}^n \langle D_i\,,\, h_i^\inv\partial_0 h_i\rangle+\int_\RR dx^0 \langle T, g\partial_0 g^\inv\rangle\nonumber\\
 &\quad+\int_\RR dx^0\int_{\surfb}\langle A_0\,,\, \tfrac{k}{2\pi}F_S -
T\delta^{(2)}(x-\xb)dx^1\wedge dx^2-\sum_{i=1}^n T_i\delta^{(2)}(x-x_{(i)})dx^1\wedge dx^2 \rangle.  \nonumber
\end{align}

Variation of the action \eqref{action} yields the  equations of motion
and constraints. Varying with respect to the Lagrange multiplier $A_0$
 we obtain the constraint
\begin{align}
\label{vara0equ} &\tfrac{k}{2\pi}F_S(x)=T\delta^{(2)}(x-\xb) dx^1\wedge dx^2+\sum_{i=1}^n T_i\delta^{(2)}(x-x_{(i)})dx^1\wedge dx^2,
\end{align}
which imposes the required delta-function singularities of the
curvature at the punctures as well as the vanishing of the
curvature $F_S$ outside the punctures. Variation  with respect to
the spatial component $A_S$ of the gauge field gives
\begin{align}
\label{varasequ} &\partial_0 A_S=d_SA_0+[A_S, A_0].
\end{align}
Combined with \eqref{vara0equ} and the decomposition
\eqref{curvdec} this equation implies  that the three-dimensional
curvature $F$ \eqref{curv} on $\RR\times \surfb$
 vanishes outside the world lines  of the punctures.
Varying   with respect to $h_i$ we obtain
\begin{align}
\label{varhiequ} &\partial_0 T_i(x^0)=[T_i(x^0), A_0(x^0,x_{(i)})],
\end{align}
and variation  of  $T$ gives
\begin{align}
\label{vartequ} &A_0(x^0,\xb)=g\partial_0 g^\inv (x^0).
\end{align}
Finally,  varying   $g$ yields
\begin{align}
\label{vargequ} &\partial_0 T=[T, g\partial_0 g^\inv].
\end{align}
Note that the evolution of both $T$ and the $T_{(i)}$ is by conjugation.
In particular, regularity is therefore preserved under evolution.

The action is invariant under  gauge transformations which combine
the usual gauge transformations of the gauge field with
transformations of the variables associated to the punctures.
Let $\gamma:\RR\times\surfb\rightarrow\prgr$   be an arbitrary
smooth function which is also well defined on the world lines of the
punctures.
Then the action is invariant under
\begin{align}
\label{astrafo} &A_S\mapsto \gamma A_S\gamma^\inv+\gamma d_S\gamma^\inv & &A_0\mapsto\gamma A_0\gamma^\inv +\gamma\partial_0\gamma^\inv &
&h_i\mapsto\gamma(x^0,x_{(i)}) h_i\\
&T\mapsto\gamma(x^0, \xb) T\gamma^\inv(x^0,\xb) & &g\mapsto \gamma(x^0,\xb) g.\nonumber
\end{align}
In addition, the  action \eqref{action} is invariant (up to a total $x^0$
derivative)  under the  transformations
\begin{align}
\label{carttrafoi} h_i\mapsto h_ic_i
\end{align}
with    functions $c_i:\RR\rightarrow \prgr$ taking values in the stabiliser
group of $D_i$, i.e. the subgroup of elements of
$\prgr$ whose coadjoint action leaves  $D_i$ invariant.
The gauge transformation \eqref{carttrafoi} arises because of
 the redundancy in the
parametrisation \eqref{tidef}  of $T_i$ via $h_i$.

For our calculation of the symplectic structure in the next section
we need
to parametrise the Lie algebra element $T$ in analogy with
\eqref{tidef}
 in terms of a general group element $h\in\prgr$
and an element $D$ in one of the Cartan subalgebras of $\pralg$.
As explained in Sect.~2, this is equivalent to assuming that $T$
is a regular element of $\pralg$. Regularity is preserved under
conjugation and, following the remark made after \eqref{vargequ},
the assumption  of regularity is thus  consistent with the
equations of motion for $T$. We assume moreover  that $T$ can be
conjugated into an abelian    Cartan subalgebra. As explained in
Sect.~2, this assumption is automatically  fulfilled for regular
elements when $G$  is semi-simple.
 We pick one such abelian Cartan subalgebra, denoted $\gothc$ as in Sect.~2,
and write $T$ as
\begin{align}
\label{tbdef} T=h D h^\inv
\end{align}
with a phase space variable  $D\in \gothc$. Note that  choices of
non-conjugate Cartan subalgebras lead, in general,  to different
theories. Also, we observe that in trading the dynamical variable
$T$ for the two dynamical variables $h$ and $D$ we have introduced
a redundancy which leads to an additional gauge invariance,
 as we shall see.

Using the parametrisation \eqref{tbdef}
we rewrite the kinetic term of the distinguished puncture in \eqref{action}
 as
\begin{align}
\label{bdterm} \langle T, g \partial_0 g^\inv\rangle=\langle D, w^\inv \partial_0 w\rangle-\langle D, h^\inv\partial_0 h \rangle
\end{align}
with
\begin{align}
\label{wdef} w=g^\inv h,
\end{align}
see also \cite{FG} where a similar coordinate transformation is
discussed. The action \eqref{action} then becomes
\begin{align}
 &S[A_S,A_0, h_i, D, h,w]=\label{action2}\\
 &\quad\;\;\int_\RR dx^0\int_{\surfb}\tfrac{k}{4\pi}\langle\partial_0 A_S\wedge A_S\rangle
 -\int_\RR dx^0 \sum_{i=1}^n \langle D_i\,,\, h_i^\inv\partial_0 h_i\rangle -\int_\RR dx^0 \langle D, h^\inv\partial_0 h\rangle\nonumber\\
 &\quad+\int_\RR dx^0\int_{\surfb}\langle A_0\,,\, \tfrac{k}{2\pi}F_S -
hDh^\inv\delta^{(2)}(x-\xb)dx^1\wedge dx^2-\sum_{i=1}^n T_i\delta^{(2)}(x-x_{(i)})dx^1\wedge dx^2 \rangle\nonumber\\
&\quad+\int_\RR dx^0 \langle D, w^\inv \partial_0 w\rangle.\nonumber
\end{align}
Expressed in the variables $h$, $w$ and $D$
the gauge transformation \eqref{astrafo} reads
\begin{align}
\label{astrafo2} &A_S\mapsto \gamma A_S\gamma^\inv+\gamma d\gamma^\inv & &A_0\mapsto\gamma A_0\gamma^\inv +\gamma\partial_0\gamma^\inv &
&h_i\mapsto\gamma(x^0,x_{(i)}) h_i\\
&h\mapsto\gamma(x^0, \xb) h & &D\mapsto D & &w\mapsto w.\nonumber
\end{align}
Furthermore, we  have the anticipated additional gauge invariance reflecting
the redundancy of the new  variables
 $D, h, w$:  given $g\in\prgr$ and
$T\in\pralg$  the relations  \eqref{tbdef} and \eqref{wdef} only
define $D\in \csg, h,w\in\prgr$ up to simultaneous
right-multiplication of $h$ and $w$ by an element
 of the stabiliser group of $D$.
As a result the action \eqref{action2} is invariant under
\begin{align}
\label{carttrafo} h\mapsto hc\qquad w\mapsto wc
\end{align}
with a  $x^0$-dependent function $c:\RR\rightarrow \prgr$ which, for
each $x^0$, takes values in the stabiliser group of $D(x^0)$.



\section{Combinatorial description of  the symplectic structure}
\label{combi}

\subsection{Outline of the method}
In this section, we derive a description of the symplectic structure on the phase space in terms of a finite number of variables
closely related to the holonomies of certain curves on the surface $\surfb$. We start with the canonical symplectic form associated
to the action \eqref{action2}
\begin{align}
\label{hwdsympform} \Omega=&-\delta\langle D, w^\inv\delta w\rangle
+\delta
\langle D, h^\inv\delta h\rangle+ \sum_{i=1}^n \delta\langle D_i,
h_i^\inv\delta h_i\rangle+\frac{k}{4\pi}\int_{\surfb} \langle \delta A_S\wedge\delta A_S\rangle,
\end{align}
where the symbol $\delta$ stands for exterior differentiation in field
space. The form \eqref{hwdsympform} is symplectic (non-degenerate) on the
enlarged phase space  parametrised by the variables $D\in\gothc$,
$D_i\in \cup_\iota \gothc_\iota$, $w,h,h_i\in\prgr$ and the spatial
gauge field $A_S$, where the conditions that the $D_i$ are fixed  and
that $A_S$ satisfies \eqref{vara0equ} are not imposed.
 Both of these conditions are first class constraints, so to obtain
the physical phase space we need  to impose the constraints {and}
divide
 the phase space by the gauge
transformations generated by these constraints. In addition,
we have to divide by the gauge transformations
\eqref{carttrafo} arising from  the redundant parametrisation
\eqref{tbdef}.
The form $\Omega$ then
determines  a unique symplectic structure on the physical phase space.

 In practice it is often
difficult or impossible to write down an explicit formula for the symplectic
structure on the physical phase space. However, in many situations
it is possible to describe the symplectic structure in terms of a two-form  $\Omega$  on an auxiliary space, in the following
referred to as extended phase space and denoted by $\mathcal{P}_{ext}$, with some (preferably most) but not all of the gauge freedom divided out.
 The remaining gauge freedom is encoded in a set of constraints defining a constraint surface $\mathcal{C}\subset  \mathcal P_{ext}$, on which
 this two-form $\Omega$ agrees with the pull-back of the symplectic structure
on the physical phase space.

For Chern-Simons theory on a spatial surface $S_{g,n}$ of genus
$g$ and with $n$ ordinary punctures, such a description has been
achieved by Alekseev and Malkin \cite{AMII}. They consider the
case of a simple, complex  gauge group $H$ with  fixed
$H$-conjugacy classes associated to each puncture. In their
description, the phase space is parametrised in terms of  a finite
set of variables linked to the holonomies along the generators of
the surface's fundamental group, and
 its symplectic structure is given in terms of its pull-back to the
 manifold $H^\ntg\times C^{n+g}$, where $C\subset H$ is a fixed Cartan subgroup.

In our derivation of the symplectic structure
defined by \eqref{hwdsympform} we closely follow the method introduced by Alekseev and Malkin \cite{AMII}. However,  we extend
their formalism to deal with the distinguished puncture at $x_\infty$ where
the curvature is not restricted to lie in a fixed conjugacy class. The
resulting extended phase space $\mathcal{P}_{ext}$ is the manifold
$(\prgr)^{n+2g+1}\times \gothc $,
where the first $\ntg$ copies of $\prgr$ correspond the generators of the surface's $\surfb$ fundamental group, the last copy stands for the
 element $w\in\prgr$ and $\gothc$ parametrises the element $D$ in \eqref{hwdsympform}.
The symplectic form $\Omega$
determines a two-form on this space which we also  denote by $\Omega$,
and for which we derive an explicit formula.

As the derivation of this two-form is both lengthy and technical,
we start by outlining the main steps in the method introduced in
\cite{AMII}. The first step is to cut the spatial surface $\surfb$
along a set of generators of its fundamental group, which results
in a simply connected polygon and $n+1$ punctured discs. The
second step is to trivialise the gauge field on the polygon and
to decompose it into a regular and a singular component on the
punctured discs.  The integral in \eqref{action2} can then be
transformed  into a set of boundary
integrals along each cut. Finally, one relates the boundary
integrals on the two sides of each cut by means of a continuity
condition on the gauge field. After evaluating and summing the two
boundary integrals in this way, one  can explicitly perform
the integration and express the result in terms of the holonomies
along the generators of the fundamental group.

\subsection{Cutting the surface}
\begin{figure}
\protect\input epsf \protect\epsfxsize=12truecm \protect
\centerline{\epsfbox{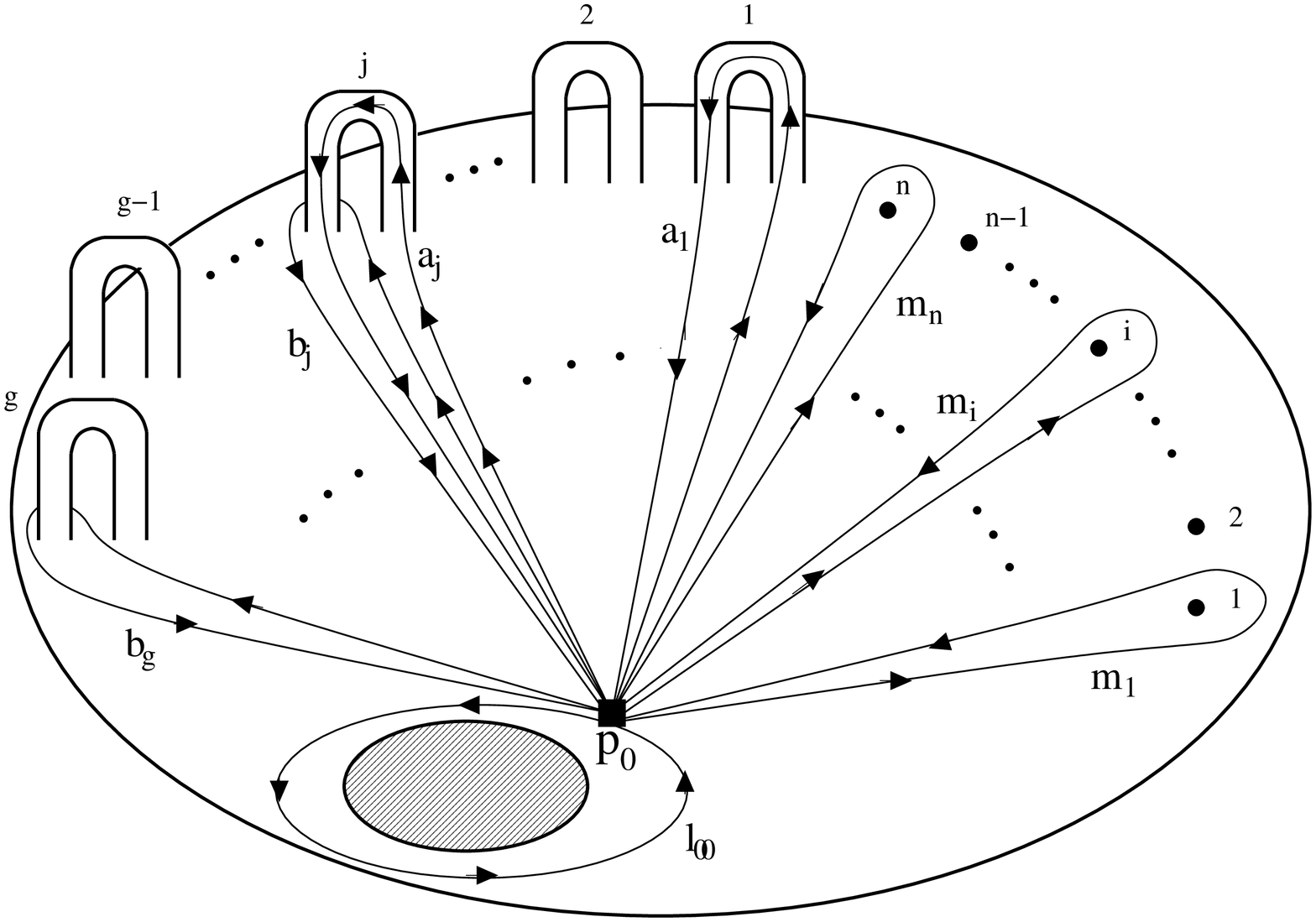}} \caption{ The generators of the
  fundamental group  $\pi_1(\surfb, \pp_0)$}
\label{pi1gen}
\end{figure}

We begin by picking a point $\pp_0$ on the surface,
distinct from the marked points $x_{(i)}$ and $x_\infty$, and
 loops
$a_j$, $b_j$, $m_i$,   $j=1,\ldots,g$,  $i=1,\ldots,n$, and $\cb$,
all based at $\pp_0$, whose homotopy classes generate   the
surface's fundamental group. There are two curves $a_j$ and $b_j$
for each handle,  one loop $m_i$ for each of the ordinary
punctures and finally one loop $\cb$ encircling the distinguished
puncture, see Fig.~\ref{pi1gen}.
The fundamental group $\pi_1(\surfb, \pp_0)$ is generated by the
homotopy classes of $a_j$, $b_j$,  $m_i$,
$j=1,\ldots,g$,$i=1,\ldots,n$, and $\cb$, subject to the relation
\bea \label{genrel} \cb
[b_n,a_n^\inv]\cdots[b_1,a_1^\inv]m_n\cdots m_1  =1. \eea We
denote the $A_S$-holonomies  along the generators by the
corresponding capital letters, so $M_i$ is the holonomy around
$m_i$, $A_j$ and $B_j$ are the holonomies around the $a$- and
$b$-cycle of the $j$-th handle, and $L_\infty $ is the holonomy
around the distinguished
 puncture.

\begin{figure}
\protect\input epsf \protect\epsfxsize=12truecm \protect
\centerline{\epsfbox{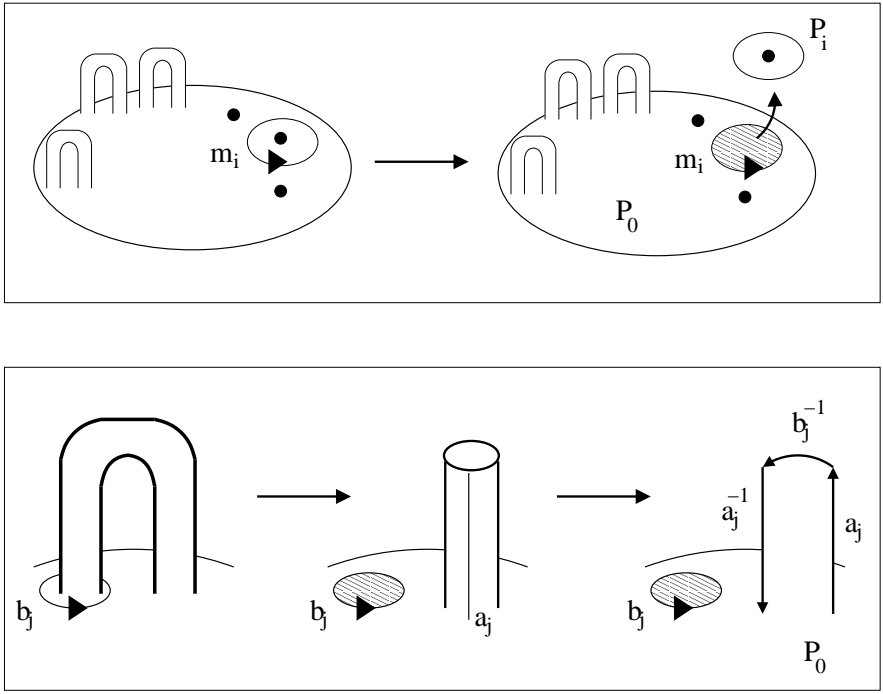}} \caption{ The cutting procedure
for punctures and handles}
\label{cut1}
\end{figure}

\begin{figure}
\protect\input epsf \protect\epsfxsize=12truecm \protect
\centerline{\epsfbox{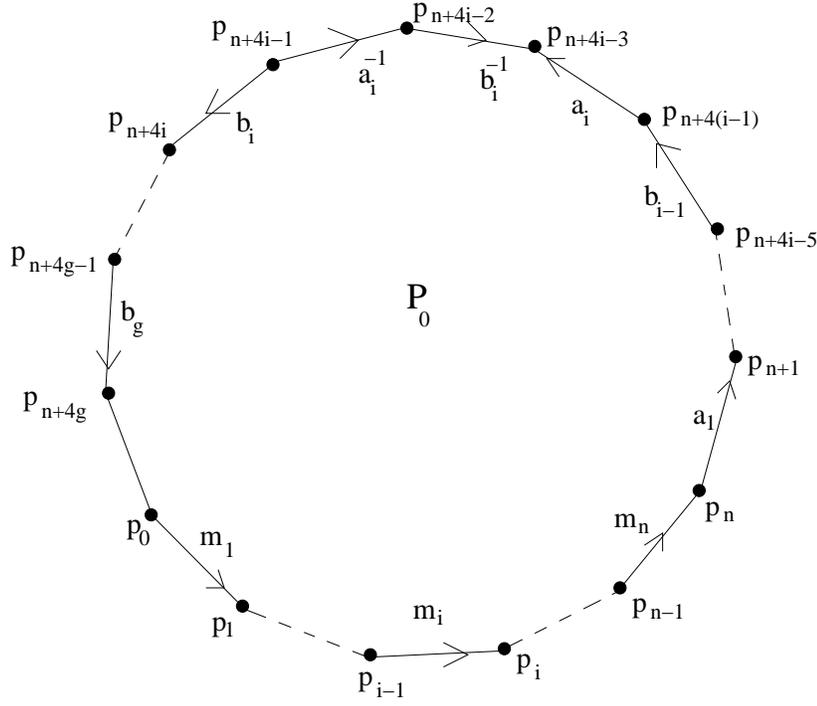}} \caption{ The polygon $P_0$} \label{cut2}
\end{figure}

The next step
is to cut the surface $\surfb$ along the curves
 $a_j$, $b_j$,  $m_i$,
$j=1,\ldots,g$, $i=1,\ldots,n$, and $\cb$,  as shown in
Fig.~\ref{cut1}. This results  in $n+2$ disconnected regions. Of
these,
 $n$ are punctured discs $P_i$ surrounding the ordinary punctures,
and one is a  punctured disc $\rb$ surrounding the distinguished
puncture. The final piece is  a  simply connected polygon $P_0$, shown
in figure
 Fig.~\ref{cut2}.

The integral in \eqref{hwdsympform} can now be written
as sum of contributions from  each  of the regions:
\bea
\label{sympformcomps}
&&\frac{k}{4\pi}\int_{\surfb} \langle \delta A_S\wedge\delta A_S\rangle
\\
&=&\frac{k}{4\pi}\int_{P_0}\langle \delta A_S\wedge\delta
A_S\rangle+\frac{k}{4\pi}\int_{\rb} \langle \delta A_S\wedge\delta
A_S\rangle+\frac{k}{4\pi}\sum_{i=1}^n \int_{P_i}\langle \delta
A_S\wedge\delta A_S\rangle. \nonumber \eea

Still following Alekseev and Malkin, we now show how to convert
each of the above integrals  into boundary integrals  and how to
express the boundary integrals in terms of $\prgr$-elements
parametrising the holonomies $M_i$, $A_j$, $B_j$  and $L_\infty $.

\subsection{Transformation into boundary integrals}

Since the interior of the polygon  $P_0$ is
 simply connected, one can express the flat gauge
field $A_S$ on $P_0$ as a pure gauge \bea \label{gamma0def2}
A_S|_{P_0}=\gamma_0d\gamma_0^\inv\qquad\text{with}\;\gamma_0(\pp_0)=1,
\eea for some $\prgr$-valued  function $\gamma_0$ on $P_0$. The
condition that $\gamma_0$ is the identity at the base point
$\pp_0$ is a partial gauge fixing. The punctured discs $P_i$,
$i=1,\ldots, n$, and $P_\infty$ are not simply connected, and
therefore it is not possible to write $A_S$ as a pure gauge here.
As explained in \cite{AMII} one can, however, introduce angular
coordinates $\phi_i$, $i=1,\ldots, n$, and $\phi$, each  with
range $[0,2\pi]$,  in the discs around the ordinary punctures and
the distinguished puncture, and give an explicit formula for a
gauge field which is flat on the punctured disc but has the
desired curvature singularity at the puncture. On each of the
discs $P_i$ such a gauge field is \bea \label{singpunct} B_i =
\tfrac{1}{k} D_i d\phi_i, \eea where $D_i$ is  the fixed element
 in $\cup_{\iota\in I}\gothc_\iota$
associated to  the $i$-the puncture. We parametrise this element explicitly
in terms of $\gothg$- and $\gothg^*$ elements  as
\bea
\label{punctpara}
 D_i=\tfrac{k}{2\pi}(\bmu_i,\bs_i).
\eea
 On the disc $P_\infty$ a flat gauge field with a curvature singularity at the
distinguished  puncture is given by\bea \label{singspunct} B=
\tfrac{1}{k} D d\phi, \eea where $D$ is  an arbitrary  element
 in the Cartan subalgebra $\csalg$, which we parametrise as
 \bea
\label{spunctpara} D=-\tfrac{k}{2\pi}(\bmu,\bs). \eea The reason
for adopting a different  sign convention here compared to
\eqref{punctpara} will be become clear later. The general form of
the gauge field on the punctured discs is obtained by applying a
gauge transformation to the singular gauge fields
\eqref{singpunct} and \eqref{singspunct}. On $P_i$ this gives \bea
\label{gammamdef2} A_S|_{P_\mi}=\tfrac{1}{k}\gamma_\mi D_i
d\phi_i\gamma_\mi^\inv+\gamma_\mi
d\gamma_\mi^\inv\quad\text{with}\quad\gamma_\mi(x_{(i)})=h_i, \eea
where the condition $\gamma_\mi(x_{(i)})=h_i$ ensures that the
constraint \eqref{vara0equ} is satisfied. Similarly, on the
punctured disc $\rb$ we can write the general gauge potential as
\bea \label{boundgf2}A_S|_{\rb}=\tfrac{1}{k}\gamma_\infty D
d\phi\gamma_\infty^\inv+\gamma_\infty d\gamma_\infty^\inv
\quad\text{with}\quad\gamma_\infty(\xb)=h. \eea

Using the  expressions \eqref{gamma0def2} and \eqref{gammamdef2}
for the gauge field, Alekseev and Malkin showed in  \cite{AMII}  how to
to transform the integrals over $P_0$ and $P_i$, $i=1,\ldots, n$
  in \eqref{sympformcomps} into boundary integrals.
The main tool is a technical lemma, summarised as Lemma
\ref{gflemma}
 in Appendix \ref{alekseevmalkin}.

Application of this lemma to the region $P_0$ gives \bea
\label{0symp2} \frac{k}{4\pi}\int_{P_0}\langle\delta
A_S\wedge\delta A_S\rangle= \frac{k}{4\pi}\int_{\partial
P_0}\langle\delta \gamma_0^\inv \gamma_0 d\left( \delta
\gamma_0^\inv \gamma_0\right)\rangle, \eea and the result for the
punctured discs $P_i$ $i=1,\ldots, n$ is \cite{AMII} \bea
\label{msymp2}
&&\frac{k}{4\pi}\int_{P_i}\langle \delta A_S\wedge\delta A_S\rangle\\
&=&-\delta \langle D_i\,,\, h_i^\inv\delta h_i
\rangle+\frac{k}{4\pi}\int_{\partial P_i} \langle \delta
\gamma_\mi^\inv \gamma_\mi d\left(\delta \gamma_\mi^\inv
\gamma_\mi \right)\rangle-\tfrac{2}{k}\delta \langle D_i\delta
\gamma_\mi^\inv \gamma_\mi\rangle d\phi_i.  \nonumber \eea To find
the corresponding result  for the punctured disc $\rb$, we use
Lemma \ref{gflemma} with $B=\tfrac{1}{k}Dd\phi$,
$\gamma=\gamma_\infty$. The term  $\langle\delta B\wedge\delta
B\rangle$  vanishes because  $D$ is in the abelian Cartan
subalgebra $\gothc$. The resulting contribution to the symplectic
structure is \bea \label{boundsymp}
&&\frac{k}{4\pi}\int_{\rb}\langle \delta A_S\wedge\delta A_S\rangle\\
&=& -\delta \langle D\,,\, h^\inv\delta h \rangle+\frac{k}{4\pi}\int_{\partial \rb} \langle \delta \gamma_\infty^\inv \gamma_\infty d\left(\delta
\gamma_\infty^\inv \gamma_\infty \right)\rangle-\tfrac{2}{k}\delta \langle  D\delta \gamma_\infty^\inv \gamma_\infty\rangle d\phi\nonumber,
\eea
which is formally the same as the contribution for the punctured discs $P_i$.
However, for further calculations
we need to keep in mind that $\delta D \neq 0$.

Collecting the expressions for the integrals over the regions
$P_0$, $P_i$, $\rb$ and inserting them into \eqref{hwdsympform},
we find that the first terms in \eqref{msymp2} and
\eqref{boundsymp} are cancelled by terms in \eqref{hwdsympform}.
Thus the two-form \eqref{hwdsympform} is given by
\begin{align}
\label{symp} \Omega =&-\delta\langle D\,,\,w^\inv\delta w\rangle+\frac{k}{4\pi}\int_{\partial \rb} \langle \delta \gamma_\infty^\inv
\gamma_\infty d\left(\delta \gamma_\infty^\inv \gamma_\infty
\right)\rangle-\tfrac{2}{k}\delta \langle  D\delta \gamma_\infty^\inv \gamma_\infty\rangle d\phi\\
+&\frac{k}{4\pi}\int_{\partial P_0}\langle\delta \gamma_0^\inv
\gamma_0 d\left( \delta \gamma_0^\inv
\gamma_0\right)\rangle+\frac{k}{4\pi}\sum_{i=1}^n\int_{\partial
P_i} \langle \delta \gamma_\mi^\inv \gamma_\mi d\left(\delta
\gamma_\mi^\inv \gamma_\mi \right)\rangle-\tfrac{2}{k}\delta
\langle  D_i\delta \gamma_\mi^\inv \gamma_\mi\rangle
d\phi_i\nonumber.
\end{align}
The
boundary integrals in \eqref{symp} involve
exactly two integrations along each cut, one from each side.
The next step in the evaluation is the summation over these two contributions by means of an overlap condition.
The gauge potential is required to be smooth, but this
requirement only determines the trivialising gauge transformation up to a
constant element in the gauge group $\prgr$. In the next subsection we explain, following \cite{AMII}, how to relate these constant elements to the holonomies
of the connection $A_S$ around the generators of the fundamental group of
 $\surfb$.

\subsection{Overlap conditions}
\label{overhol}
The boundary of the polygon $P_0$ has the decomposition
\begin{align}
\label{polysides} \partial P_0=\cb \;\cup\;\bigcup_{i=1}^n\; m_i
\bigcup_{i=1}^g \left( a_i \cup b_i\cup a_i^\inv\cup
b_i^\inv\right).
\end{align}

For the cuts along the generators $m_1,\ldots,m_n$  the continuity
of the gauge field amounts to the condition \bea \label{moverl}
\gamma_0d\gamma_0^\inv|_{m_i} =\left(\tfrac{1}{k}\gamma_\mi D_i
d\phi_i\gamma_\mi^\inv+\gamma_\mi d\gamma_\mi^\inv\right)|_{m_i},
\eea which can be expressed equivalently as \bea \label{mover}
\gamma_0^\inv|_{m_i}=N_\mi
D_\mi(\phi_i)\gamma_\mi^\inv|_{m_i}\quad\text{with}\quad dN_\mi=0,
\eea
 where $C_\mi(\phi_i)=\exp(\tfrac{1}{k}D_i\phi_i)$
 and $N_\mi$ is an arbitrary but  constant element of the gauge group
$\prgr$.
For the cuts along $a_1,b_1,\ldots, a_g,b_g$ the continuity condition is
 \bea
\label{hoverl}
\gamma'_0d\left(\gamma'_0\right)^{-1}|_x=\gamma''_0
d\left(\gamma''_0\right)^{-1}|_x,
\eea
where $x=a_1,b_1,\ldots,a_g,b_g$ and  $\gamma'_0,\gamma''_0$
denote the values of the
trivialising gauge transformation $\gamma_0$
at the two sides of the polygon $P_0$ associated to the curve $x$.
The condition \eqref{hoverl} is equivalent to
\bea
\label{hover}
\left(\gamma'_0\right)^{-1}|_x=N_x \lbr \gamma''_0 \rbr^{-1}|_x\quad\text{with}\quad dN_x=0.
\eea
Finally, for the cut along the curve $\cb$ we have
\bea
\label{boundoverl}
\gamma_0d\gamma_0^\inv|_{\cb}
=\left(\tfrac{1}{k}\gamma_\infty D d\phi \gamma_\infty^\inv+\gamma_\infty
d\gamma_\infty^\inv\right)|_{\cb}
\eea
or, equivalently,
\bea
\label{boundover}
\gamma_0^\inv|_{\cb}=N_\infty C_\infty(\phi)
\gamma_\infty^\inv|_{\cb}\quad\text{with}\quad dN_\infty=0,
\eea
where $C_\infty(\phi)=\exp(\frac{1}{k} D\phi)$.

 The values of  $\gamma_0$, $\gamma_\mi$ and $ \gamma_\infty$ at the
endpoints of the cuts are related to the holonomies of the
generators of $\pi_1(\surfb, \pp_0)$. We start by considering the
polygon $P_0$.  Denoting the endpoints of the cuts by $\pp_i$,
$i=0,\ldots,n+4g$, as shown in Fig.~\ref{cut1}, so that in
particular $m_i$ runs from $\pp_{i-1}$ to $\pp_i$ and $\cb $ from
$\pp_{n+4g}$ to $\pp_0$, one finds that the  parallel transport
along the cut with endpoints  $\pp_{i-1}$ and $\pp_i$ is given by
\bea \label{holrule} PT_{\pp_{i-1}\rightarrow \pp_i}=
\gamma_0(\pp_i)\gamma_0^\inv(\pp_{i-1}). \eea As mentioned after
\eqref{gamma0def2}, one can set $\gamma_0(\pp_0)=1 $, thus
partially  fixing the gauge. Then the values of $\gamma_0$ at the
endpoints of the cuts are related to the holonomies $A_j$, $B_j$,
$M_i$, $j=1,\ldots,g$, $i=1,\ldots,n$,
 and $L_\infty$
by the following equations:
\begin{align}
\label{cornerval2}
&\gamma_0(\pp_i)=K_i^\inv:=M_i\cdots M_1\qquad\text{for}\;1\leq i\leq n\\
&\gamma_0(\pp_{n+4j-3})=K_{n+2j-1}^\inv:=A_j[B_{j-1},A_{j-1}^\inv]\cdots[B_1,A_1^\inv]M_n\cdots M_1\nonumber\\
&\gamma_0(\pp_{n+4j-2})=B_j^\inv A_j[B_{j-1},A_{j-1}^\inv]\cdots[B_1,A_1^\inv]M_n\cdots M_1\nonumber\\
&\gamma_0(\pp_{n+4j-1})=A_j^\inv B_j^\inv A_j[B_{j-1},A_{j-1}^\inv]\cdots[B_1,A_1^\inv]M_n\cdots M_1\nonumber\\
&\gamma_0(\pp_{n+4j})=K_{n+2j}^\inv:=[B_j,\aj^\inv]\cdots[B_1,A_1^\inv]M_n\cdots M_1\quad\text{for}\; 1\leq j\leq g\nonumber.
\end{align}
Alternatively, considering the parallel transport from $\pp_0$ to $\pp_{n+4g}$
 we can write the condition for $\pp_{n+4g}$ in terms of
 $L_\infty$, using \eqref{holrule} and $\gamma(\pp_0)=1$:
\bea \gamma_0(\pp_{n+4g})=L_\infty^{-1}, \eea so that - in
agreement with \eqref{genrel} - the holonomy $L_\infty$ is given
by \bea \label{genrell} L_\infty
=\left([B_n,A_n^\inv]\cdots[B_1,A_1^\inv]M_n\cdots
M_1\right)^\inv. \eea

Note that the overlap conditions \eqref{hoverl} ensure that the holonomies along the two sides of $P_0$ corresponding to each $a$- and $b$-cycle are indeed the same
 \begin{align}
\label{handhol02}
&A_i=\gamma_0(\pp_{n+4i-3})\gamma_0^\inv(\pp_{n+4(i-1)})=\gamma_0(\pp_{n+4i-2})\gamma_0^\inv(\pp_{n+4i-1})\\
&B_i=\gamma_0(\pp_{n+4i})\gamma^\inv_0(\pp_{n+4i-1})=\gamma_0(\pp_{n+4i-3})\gamma_0^\inv(\pp_{n+4i-2}).\nonumber
\end{align}

We now consider the punctured discs $P_i$ for the ordinary punctures.
The parallel transport around $P_i$ gives the holonomy $M_i$, which we can now
relate to
the parametrisation of the gauge field $A_S|_{m_i}$ in
\eqref{gammamdef2}. The gauge transformation $\gamma_\mi$ is single-valued, so
$\gamma_\mi(\pp_{i-1})=\gamma_\mi(\pp_i)$ and therefore
\begin{align}
\label{dischol2} M_i=g_i C_i^\inv g_i^\inv\qquad\text{with}\qquad
C_i=\exp(\tfrac{2\pi}{k} D_i)= (e^{\bmu_i},
\bs_i),\;g_i=\gamma_\mi(\pp_i).
\end{align}
Similarly, on the  punctured disc $\rb$ we have the expression \eqref{boundgf2}
with $\gamma_\infty(\pp_0)=\gamma_\infty(\pp_{n+4g})$.
 Recalling  the  parametrisation of $D$ in \eqref{spunctpara} we have
\bea
\label{boundholgamma}
L_\infty=K_\ntg=g_\infty
C^\inv g_\infty^\inv, \quad
\text{with}\quad C=\exp(\tfrac{2\pi}{k} D)=
(e^{-\bmu},-\bs),\;g_\infty=\gamma_\infty(\pp_0).
\eea

\subsection{Evaluation in terms of holonomies}

\label{evalsect}
We will now combine the two integrals along each cut and express the two-form \eqref{symp} in terms of the holonomies $\mi,\aj,\bjj, L_\infty$.
We begin with
the contributions of all cuts except the one around the distinguished
puncture:
\begin{align}
\Omega_b:=&\frac{k}{4\pi}\int_{\partial P_0-\cb}\langle\delta
\gamma_0^\inv \gamma_0 d\left( \delta \gamma_0^\inv
\gamma_0\right)\rangle \\
+&\frac{k}{4\pi}\sum_{i=1}^n\int_{\partial P_i} \langle \delta
\gamma_\mi^\inv \gamma_\mi d\left(\delta \gamma_\mi^\inv
\gamma_\mi \right)\rangle-\tfrac{2}{k}\delta \langle  D_i\delta
\gamma_\mi^\inv \gamma_\mi\rangle d\phi_i.\nonumber \end{align}
For the case where the gauge group $H$ is a simple complex group
or its compact  real form a  formula for $\Omega_b$ was given by
Alekseev and Malkin in \cite{AMII} as
 a sum of the contributions $\Omega_\mi$,
 $i=1,\ldots, n$,
 from the ordinary punctures \eqref{mcontrib} and the
contributions $\Omega_{H_j}$, $i=1,\ldots, g$, from the handles
\eqref{hcontribcsa}. In Appendix \ref{alekseevmalkin} we quote
their expressions for $\Omega_\mi$ and $\Omega_{H_j}$,  sketch
their derivation, and also explain modifications  required for our
gauge groups $\prgr$, see Lemma \ref{amlemma3}, Lemma
\ref{amlemma4} and Lemma \ref{ourcase}. Summing over the
contributions $\Omega_\mi$, $\Omega_{H_j}$ for each ordinary
puncture and handle, one finds
\begin{align}
\label{modform2}
\Omega_b=&\sum_{i=1}^n
\Omega_\mi+\sum_{i=1}^g \Omega_{H_i}\nonumber\\
=&\frac{k}{4\pi}\sum_{i=1}^{n} \langle C_i g_i^\inv \delta g_i
C_i^\inv\wedge g_i^\inv\delta g_i\rangle-\frac{k}{4\pi}\sum_{i=1}^\ntg
\langle \delta K_i K_i^\inv\wedge \delta K_{i-1} K_{i-1}^\inv\rangle\\
-&\frac{k}{4\pi}\sum_{i=1}^{g}
\left(\langle A_i^\inv \delta A_i\wedge B_i^\inv \delta B_i\rangle +
\langle \delta(B_iA_iB_i^\inv)B_iA_i^\inv B_i^\inv \wedge \delta B_i\,
B_i^\inv
\rangle\right). \nonumber
\end{align}

It remains to evaluate the contribution to the two-form \eqref{symp} from the cut $\cb$ around the distinguished puncture, which is given by
\bea
\label{boundcont}
\Omb:&=&-\delta\langle D\,,\,w^\inv\delta w\rangle
+ \frac{k}{4\pi}\int_{\partial P_0\cap\partial P_\infty}\langle
\delta\gamma_0^\inv\gamma_0d(\delta\gamma_0^\inv\gamma_0) \rangle \\
&&+\frac{k}{4\pi}\int_{\partial \rb} \langle \delta \gamma_\infty^\inv \gamma_\infty d\left(\delta
\gamma_\infty^\inv \gamma_\infty \right)\rangle-\tfrac{2}{k}\delta
\langle  D\delta \gamma_\infty^\inv \gamma_\infty\rangle d\phi.
\nonumber
\eea
The overlap condition \eqref{boundover} implies
\begin{align}
\label{overlrel}
\langle \delta \gamma_0^\inv \gamma_0 d(\delta \gamma_0^\inv \gamma_0)\rangle=&\langle \delta \gamma_\infty^\inv \gamma_\infty d(\delta \gamma_\infty^\inv  \gamma_\infty)\rangle-2\delta\langle C_\infty^\inv d C_\infty \delta \gamma_\infty^\inv \gamma_\infty \rangle\\
+&d\langle \delta N_\infty N_\infty^\inv \delta\gamma_0^\inv \gamma_0\rangle+d \langle C_\infty^\inv\delta C_\infty
\delta\gamma_\infty^\inv\gamma_\infty\rangle,\nonumber
\end{align}
and we obtain \bea \label{boundcontrib} \Omb =-\delta \langle
D,w^\inv\delta w\rangle+\tfrac{k}{4\pi} \langle\delta N_\infty
N_\infty^\inv\delta \gamma_0^\inv\gamma_0
\rangle|_{\pp_{n+4g}}^{\pp_0}+\tfrac{k}{4\pi}\langle
C_\infty^\inv\delta C_\infty,
\delta\gamma_\infty^\inv\gamma_\infty\rangle|_{\pp_{n+4g}}^{\pp_0}.
\eea Recalling the notation $g_\infty
=\gamma_\infty(\pp_0)=\gamma_\infty(\pp_{n+4g})$ from
\eqref{boundholgamma} and $ C_\infty(\phi)=\exp(\frac{1}{k}D\phi)$
\eqref{boundover}, we have for the last term in
\eqref{boundcontrib}
\begin{align}
\label{term3bcontrib}
 \tfrac{k}{4\pi}\langle C_\infty^\inv\delta C_\infty,
\delta\gamma_\infty^\inv\gamma_\infty\rangle|_{\pp_{n+4g}}^{\pp_0}=
&-\tfrac{1}{4\pi}(\phi(\pp_0)-\phi(\pp_{n+4g}))\langle\delta
D, g_\infty^\inv\delta g _\infty \rangle\\
=&-\tfrac{1}{2}\langle\delta D\,,\,g_\infty^\inv\delta g_\infty
\rangle \nonumber \\
=&\tfrac{k}{4\pi}\langle (\delta\bmu,
\delta\bs)\,,\, g_\infty^\inv\delta g_\infty\rangle,\nonumber
\end{align}
where the last equality follows from the definition of $C$ in
\eqref{boundholgamma} and we used
$\delta\phi(\pp_0)=\delta\phi(\pp_{n+4g})=0$,
$\phi(\pp_0)-\phi(\pp_{n+4g})=2\pi$.
To evaluate the second term in \eqref{boundcontrib},
we recall that $\gamma_0(\pp_{n+4g})=K_{n+2g}^\inv$ and that
we imposed the condition $\gamma_0(\pp_0)=1$.
Together with \eqref{boundover}, this
implies $N_\infty=g_\infty e^{-\frac{1}{k}D\phi(\pp_0)}$
and, using \eqref{boundholgamma}, we find
\begin{align}
\label{boundcontrib2} \Omb=
\tfrac{k}{2\pi}\delta\langle (\bmu,\bs), w^\inv\delta w\rangle+
\tfrac{k}{4\pi}\langle K_\ntg^\inv \delta
K_\ntg, \delta g_\infty g_\infty^\inv \rangle+\tfrac{k}{4\pi}\langle
(\delta\bmu,\delta\bs)\,,\,
g_\infty^\inv\delta g_\infty\rangle.
\end{align}

Adding this expression to the contribution \eqref{modform2} from the cuts along the generators of the fundamental group, we find that the two-form
\eqref{symp} is given by
\begin{align}
\label{intermsymp}
\Omega=&\frac{k}{4\pi}\sum_{i=1}^{n} \langle C_i g_i^\inv \delta g_i
C_i^\inv\wedge g_i^\inv\delta g_i\rangle-\frac{k}{4\pi}\sum_{i=1}^\ntg
\langle \delta K_i K_i^\inv\wedge \delta K_{i-1} K_{i-1}^\inv\rangle\\
-&\frac{k}{4\pi}\sum_{i=1}^{g}\left(
\langle A_i^\inv \delta A_i\wedge B_i^\inv \delta B_i\rangle +
\langle \delta(B_iA_iB_i^\inv)B_iA_i^\inv B_i^\inv \wedge \delta B_i\,
B_i^\inv
\rangle \right)\nonumber \\ \
+&\frac{k}{2\pi}\delta\langle (\bmu,\bs), w^\inv\delta w\rangle+
\frac{k}{4\pi}\langle  K_\ntg^\inv \delta K_\ntg,\delta g_\infty g_\infty^\inv
\rangle+\frac{k}{4\pi}\langle
(\delta\bmu,\delta\bs)\,,g_\infty^\inv
\delta g_\infty\rangle.\nonumber
\end{align}

\section{The symplectic structure for gauge groups  $\prgr$ }

Equation \eqref{intermsymp} is one of the main results of this
paper. It gives an expression for the symplectic structure on the
phase space in terms of the pull-back to the group
$(\prgr)^{n+2g+2}\times \gothc$ in terms of the variables $g_1,
\ldots,g_n$, $A_1,B_2, \ldots,A_g,B_ g$,
 $g_\infty$, $w$ and $(\bmu,\bs)$ and the constant elements
 $C_1,\ldots, C_n$.
Although we have written this result using the explicit
parametrisation of $D$ in terms of the $\gothg$-element
$\bmu$ and the $\gothg^*$-element $\bs$, the result itself does not depend
this
parametrisation. When replacing $ (\bmu,\bs)$ by $-\frac{2\pi}{ k} D$
the formula \eqref{intermsymp} holds for any Lie group $H$ with an
invariant pairing $\langle \, , \,\rangle$ on its Lie algebra.

In this section, we derive the other central result of our paper,
the explicit evaluation of the two-form \eqref{intermsymp} for
gauge groups of the form $\prgr$. The calculations in this section
make repeated use of two parametrisations of group elements in
$\prgr $. First, we write elements of $\prgr $ in terms of an
element $u\in G$ and an element $\bj \in \gothg^*$ as
$(u,-\Ad^*(u^\inv)\bj)$. As explained in Appendix
\ref{heisendouble},  this parametrisation is derived from a
diffeormophism between $\prgr$ and the dual Poisson-Lie group
$G\times \gothg^*$. The second parametrisation has its origins in
dressing actions of $\prgr$. As also explained in
\ref{heisendouble} the dressing action of $(v,\bx)\in \prgr$ on an
element $(\tilde u,- \tilde \pbj) \in G\times \gothg^*$ is simply
conjugation of the corresponding element $(\tilde u,- \Ad^*(\tilde
u^\inv) \tilde \pbj)$ in $\prgr$. If $(\tilde u,- \tilde \pbj)
=(e^{-\bmu},-\bs)$ is in the abelian subgroup $T_\gothc$ we obtain
the following  parametrisation of elements in its  dressing orbit:
\bea \label{orbitparam} (u,-\Ad^*(u^\inv)\bj)=(v,\bx)(e^{-\bmu},
-\bs)(v,\bx)^{-1}. \eea The resulting expressions for $u$ and
$\bj$ in terms of $(v,\bx)$ and
 $(e^{-\bmu},-\bs)$ are given in \eqref{ujform}.

We begin with  a technical lemma concerning the parametrisation
\eqref{orbitparam}
which  can be proved by direct calculation. Equipped with this lemma we then show how to write the two-form \eqref{intermsymp}
as the exterior derivative of an explicitly given one-form.
\begin{lemma}
\label{lemma1} For $K^\inv=g C^\inv
g^\inv=(u,-\Ad^*(u^\inv)\bj)\in\prgr$
with  general elements $g=(v,\bx)\in\prgr$
and  elements  $C=(e^\bmu, \bs)$ in the abelian  subgroup $T_\gothc$,
we have the following identity
\begin{align}
 \label{helpid1}
&\langle C g^\inv\delta g C^\inv\wedge g^\inv\delta g\rangle-2\langle\delta
CC^\inv \,,\,g^\inv\delta g\rangle=- \langle
K^\inv  \delta K\,, \delta g g^\inv  \rangle-
\langle (\delta \bmu, \delta \bs)\,,\,
g^\inv\delta g\rangle\nonumber\\
&= - \langle\delta \bj\,,\, u^\inv\delta u\rangle -
\langle \delta \bs\,,\, \delta \bmu \rangle - 2\delta\langle \bj\,,\,\delta v
v^\inv\rangle+2\delta\langle\bx\,,\, v \delta\bmu v^\inv \rangle.\end{align}
\end{lemma}

\begin{theorem}
\label{prgrtheorem}
If the holonomies around
 the generators of $\pi_1(\surfb,\pp_0)$  are parametrised as
\begin{align}
\label{holparam}
&\mi=(u_\mi,-\Ad^*(u_\mi^\inv)\bj_\mi)=(v_\mi,\bx_\mi)(e^{-\bmu_i},-\bs_i)(v_\mi,\bx_\mi)^\inv\\
&\ai=(u_\ai,-\Ad^*(u_\ai^\inv)\bj_\ai)\nonumber\\
&\bi=(u_\bi,-\Ad^*(u_\bi^\inv)\bj_\bi)\nonumber,
\end{align}
with the  inverse  holonomy of the curve $\cb$
around the distinguished puncture written as
\begin{align} \label{jtotal} L_\infty^\inv=&K_\ntg^\inv=(u_{K_\ntg},
-\Ad^*(u^\inv_{K_\ntg})\bj_{K_{\ntg}})=[B_g,A_g^\inv]\cdots[B_1, A_1^\inv] M_n\cdots M_1\\
=&(v_\infty,\bx_\infty)(e^{-\bmu},-\bs)(v_\infty,\bx_\infty)^\inv\nonumber
\end{align} the two-form \eqref{intermsymp} is given by
\bea
\label{gravom} \Omega= \frac{k}{2\pi}\delta \left(\langle (\bmu,\bs),
  w^\inv\delta w\rangle+\Theta+ \langle \bj_{K_\ntg},\delta v_\infty
v_\infty^\inv\rangle -\langle
\Ad^* (v_\infty) \bx_\infty-\tfrac{1}{2}\bs \,,\,\delta
\bmu\rangle\right)
\eea
with the one-form
\begin{align}
\label{gravtheta}
\Theta=&\sum_{i=1}^n \langle\; \delta(u_{M_{i-1}}\cdots u_{M_1})(u_{M_{i-1}}\cdots u_{M_1})^\inv- \delta v_\mi v_\mi^{-1}\,,\,
\bj_\mi\rangle\\
+&\sum_{i=1}^g\langle\; \delta(u_{H_{i-1}}\cdots u_{M_1})(u_{H_{i-1}}\cdots u_{M_1})^\inv\,,\, \bj_\ai\rangle\nonumber\\
&\qquad-\langle\; \delta(u_\ai^\inv u_\bi^\inv u_\ai u_{H_{i-1}}\cdots u_{M_1})(u_\ai^\inv u_\bi^\inv u_\ai u_{H_{i-1}}\cdots u_{M_1})^\inv\,,\,
\bj_\ai \rangle\nonumber\\
+&\sum_{i=1}^g\langle\; \delta(u_\ai^\inv u_\bi^\inv u_\ai u_{H_{i-1}}\cdots u_{M_1})(u_\ai^\inv u_\bi^\inv u_\ai u_{H_{i-1}}\cdots
u_{M_1})^\inv\,,\,
\bj_\bi \rangle\nonumber\\
&\qquad-\langle\; \delta( u_\bi^\inv u_\ai u_{H_{i-1}}\cdots u_{M_1})( u_\bi^\inv u_\ai u_{H_{i-1}}\cdots u_{M_1})^\inv\,,\, \bj_\bi
\rangle ,\nonumber\\
\nonumber\\
&\text{where} \quad
u_{H_i}=[u_{\bi},u_\ai^\inv]=u_\bi u_\ai^\inv u_\bi^\inv u_\ai.\nonumber
\end{align}

\end{theorem}

{\bf Proof:}

The proof is a rather lengthy calculation. We outline the main steps and some auxiliary identities used in the calculations.

1. To evaluate the sum $\sum_{i=1}^\ntg \langle \delta K_i K_i^\inv\wedge \delta K_{i-1} K_{i-1}^\inv\rangle$,  one recalls that the group
elements $K_i$ are given by
\begin{align}
\label{kidef} &K_i=M_1^\inv\cdots M_i^\inv\qquad 1\leq i\leq n\\
&K_{n+2i-1}=M_1^\inv\cdots M_n^\inv [B_1,A_1^\inv]^\inv\cdots [B_{i-1}, A_{i-1}^\inv]^\inv A_i^\inv\nonumber\\
&K_{n+2i}=M_1^\inv\cdots M_n^\inv [B_1,A_1^\inv]^\inv\cdots [B_{i}, A_{i}^\inv]^\inv\qquad\qquad 1\leq i\leq g\nonumber.
\end{align}
Inserting the parametrisation \eqref{holparam} of the holonomies along the generators
of the fundamental group, one can express these group elements as functions of the variables $u_\mi, \bj_\mi$, $u_\aj,\bj_\aj$ and $u_\bjj,\bj_\bjj$
in \eqref{holparam}
\begin{align}
\label{kidefuj}&K_i^\inv=:(u_{K_i},-\Ad^*(u_{K_i}^\inv)\bj_{K_i})\qquad
1\leq i \leq n+2g
\end{align}
with
\begin{align}
\label{ukidef}
&u_{K_i}=\begin{cases} u_{\mi}\cdots u_{\me} & 1\leq i\leq n\\
u_\aj u_{H_{j-1}}\cdots u_{H_1} u_{\mf}\cdots u_{\me} & i=n+2j-1\\
 u_{H_{j}}\cdots u_{H_1} u_{\mf}\cdots u_{\me} & i=n+2j
\end{cases}\\
\intertext{}
\label{jkidef}
&\bj_{K_{i}}=\sum_{j=1}^i\Ad^*( u_{M_{j-1}}...u_\me)\bj_{M_j}\qquad\qquad 1\leq
i\leq n\\
&\bj_{K_{n+2i-1}}=\Ad^*(u_{H_{i-1}}...u_\me)\bj_\ai+\sum_{k=1}^n\Ad^*(u_{M_{k-1}}...u_\me)\bj_{M_k}+\sum_{k=1}^{i-1}\Ad^*(u_{H_{k-1}}...u_\me)\bj_{H_k}\nonumber\\
&\bj_{K_{n+2i}}=\Ad^*(u_{H_{i-1}}...u_\me)\bj_{H_i}+\sum_{k=1}^n\Ad^*(u_{M_{k-1}}...u_\me)\bj_{M_k}+\sum_{k=1}^{i-1}\Ad^*(u_{H_{k-1}}...u_\me)\bj_{H_k}
,\nonumber\\
\intertext{}
\label{grjothi}
&u_{H_i}:=u_\bi u_\ai^{-1}u_\bi^{-1}u_\ai\hfill\\
&\bj^{H_i}:=( 1-\Ad^*(u_\ai^\inv u_\bi^\inv u_\ai))\bj^\ai+(\Ad^*(u_\ai^{-1}u_\bi^\inv
u_\ai)-\Ad^*(u_\bi^{-1} u_\ai))\bj^\bi.\nonumber
\end{align}
One then computes the derivatives $\delta K_i K_i^\inv$ using the identities
\begin{align}
\label{helpid} &(v,\bx)^\inv\delta(v,\bx)=(v^\inv\delta v,
\Ad^*(v)\delta\bx)\qquad\forall v\in G, \bx\in\mathfrak{g}^*\\
&\delta(\Ad^*(v^\inv)z)=\Ad^*(v^\inv)\delta z+\Ad^*(v^\inv)[v^\inv \delta v, z]\qquad\forall v\in G,z\in\gothg^*.\nonumber
\end{align}
 Taking into account that the bilinear form
$\langle\,,\,\rangle$ pairs the $\gothg$-component of the first argument with the $\mathfrak{g}^*$-component of the second and vice
 versa, one obtains
\begin{align}
\label{kres} &\sum_{i=1}^{\ntg}\!\langle \delta K_i K_i^\inv\!\wedge\!\delta K_{i-1}K_{i-1}^\inv\rangle\!=\!\langle \delta
\bj_{K_{n+2g}},u_{K_{n+2g}}^\inv\!\!\!\delta u_{K_{n+2g}}\rangle\!-\!\sum_{i=1}^n\langle\delta
\bj_\mi  , u_\mi^\inv\delta u_\mi\rangle\\
&-\sum_{i=1}^g\!\langle \delta \bj_\ai ,  u_\ai^\inv\delta
u_\ai\rangle\!+\!\langle \delta\left(\Ad^*(u_\ai^\inv
u_\bi^\inv)(\bj_\bi\!\!-\!\!\bj_\ai)\!-\!\Ad^*(u_\bi^\inv)\bj_\bi\right),
u_\bi u_\ai
u_\bi^\inv\delta(u_\bi u_\ai^\inv u_\bi^\inv) \rangle\nonumber\\
&-2\sum_{i=1}^n \delta\langle
  \bj_{M_i}, \delta(u_{M_{i-1}}\cdots u_{M_1} )u_{M_1}^\inv\cdots
  u_{M_{i-1}}^\inv\rangle-2\sum_{i=1}^{2g}\delta\langle \bj_{K_{n+i}}, \delta u_{K_{n+i-1}}
  u_{K_{n+i-1}}^\inv \rangle\nonumber.
\end{align}

2. For the terms containing elements $C_i$, $i=1,\ldots,n$ one
uses identities \eqref{dischol2} which express the holonomies
$M_i$ around the punctures   in terms of the variables $g_i$ and
$C_i$ together with
 Lemma \ref{lemma1}.
Taking into account that $\delta C_i=0$ for $i=1,\ldots,n$ and
 we find that the terms of the
form $\langle\delta\bj , u^\inv\delta u\rangle$ in \eqref{kres}
and \eqref{helpid1} cancel.
 To evaluate the terms involving
$A_j, B_j$, $j=1,\ldots, g$, one inserts the parametrisation
\eqref{holparam} of the holonomies $\aj,\bjj$ into the second line
of \eqref{intermsymp} and finds
\begin{align}
\label{abterm} &-\langle A_i^\inv \delta A_i\wedge B_i^\inv \delta
B_i\rangle -\langle \delta(B_iA_iB_i^\inv)B_iA_i^\inv B_i^\inv
\wedge \delta B_i\, B_i^\inv \rangle=\\
&\qquad\qquad\langle \delta\left(\,
\Ad^*(u_\bi^\inv)(\bj_\bi-\bj_\ai)-\Ad^*(u_\ai
u_\bi^\inv)\bj_\bi\,\right)\,,\, \delta(u_\bi u_\ai^\inv
u_\bi^\inv)u_\bi u_\ai u_\bi^\inv\rangle\nonumber\\
&\qquad\qquad-2\delta\langle\bj_\bi-\bj_\ai\,,\, u_\bi^\inv\delta
u_\bi\rangle+2\delta\langle\bj_\bi\,,\, u_\ai
u_\bi^\inv\delta(u_\bi u_\ai^\inv)\rangle-\langle\delta \bj_\ai
\,,\, u_\ai^\inv\delta u_\ai\rangle\nonumber.
\end{align}
Adding this term to the second line in \eqref{kres} and then using
the definition \eqref{jkidef} to evaluate the last term in
\eqref{kres} then yields \begin{align} \label{intermsymp2}
\Omega=&\tfrac{k}{2\pi}\delta\langle (\bmu,\bs), w^\inv\delta
w\rangle+\tfrac{k}{2\pi}\delta \Theta-\tfrac{k}{4\pi}\langle
\delta \bj_{K_\ntg}\,,\,u_{K_\ntg}^\inv\delta
u_{K_\ntg}\rangle\\
+&\tfrac{k}{4\pi}\langle K_\ntg^\inv  \delta K_\ntg,
\delta g_\infty g_\infty^\inv  \rangle+\tfrac{k}{4\pi}\langle
(\delta\bmu,\delta\bs)\,,\,g_\infty^\inv\delta g_\infty \rangle.\nonumber
\end{align}

3. To simplify the remaining terms, we use identity
\eqref{boundholgamma} for the holonomy around the distinguished
puncture  and, again,  Lemma \ref{lemma1}, this time with
$K=K_\ntg$, $g=g_\infty=(v_\infty,\bx_\infty)$,
$C=\exp(\frac{2\pi}{k}D)=(e^{-\bmu},-\bs)$. This allows one to
eliminate the last three terms in \eqref{intermsymp2} and finally
gives
\begin{align}
\label{intermsymp3} \Omega=&\tfrac{k}{2\pi}\,\delta\langle
(\bmu,\bs), w^\inv\delta w\rangle+\tfrac{k}{2\pi}\,\delta
\Theta+\tfrac{k}{4\pi}\langle\delta\bs,\delta
\bmu\rangle+\tfrac{k}{2\pi}\,\delta\langle \bj_{K_\ntg},\delta
v_\infty v_\infty^\inv\rangle\\
-&\tfrac{k}{2\pi}\,\delta\langle \Ad^*(v_\infty)\bx_\infty,\delta
\bmu\rangle.\qquad\qquad\qquad\qquad\qquad\qquad\qquad\qquad\qquad\qquad\Box\nonumber
\end{align}

The expression \eqref{intermsymp3}, or equivalently \eqref{gravom},
is  an explicit formula for the
pull-back of the symplectic form on the moduli space to the extended phase
space. The pull-back is the exterior derivative of a sum of one-forms which
deserve further comment.
As explained in  Appendix~\ref{heisenberg}, the first term in
\eqref{intermsymp3} is a reduction of the symplectic structure on the
contangent bundle of  $\prgr$ while
the last two terms in the expression \eqref{intermsymp3}
are  a reduction of
the canonical symplectic structure on the Heisenberg double of
$\prgr$, see the discussion preceding \eqref{favouritepot}.
The symplectic potential $\Theta$ contains the contributions from
the handles and the ordinary
punctures. It was first given in \cite{we2} in the context of
an investigation
of  Poisson structures arising in
Chern-Simons theory with gauge group $\prgr$ on
$\RR \times \surf$, where  $\surf$ is a surface of genus $g$ and with $n$
ordinary punctures. Applying results from   \cite{AMII}
it was shown there that in suitable ``decoupled'' coordinates
 $\Theta$ is a sum of standard  symplectic potentials, namely a copy
of the symplectic potential of the Heisenberg double \eqref{favouritepot}
 for every handle,
and a copy of the pull-back of the symplectic potential on the conjugacy
classes of $\prgr$. As explained after \eqref{favouritepot} the latter are
 the symplectic leaves of the dual Poisson-Lie group $G\times \gothg^*$.

The structural  relationship  between the two-form
\eqref{intermsymp3} and the Heisenberg double (as well as the
various symplectic forms derived from it) generalises to arbitrary
Poisson-Lie groups. However, the explicit and rather simple
formula we were able to give for the two-form $\Omega$ is a
reflection of the relatively simple structure of   Poisson-Lie
groups of the form $\prgr$. We will point out some uses of the
formula \eqref{intermsymp3} the Chern-Simons formulation of
(2+1)-dimensional gravity in our conclusion.

\section{Gauge invariance,  basepoint independence and gauge fixing}

Theorem \ref{prgrtheorem} gives an expression for the  two-form
\eqref{gravom} in
terms of the $\prgr$ elements parametrising the
 holonomies with respect to given basepoint $\pp_0$ and the variables $w\in\prgr$, $(e^{-\bmu}, -\bs) \in T_\gothc$. As explained
at the beginning of Sect.~\ref{combi},
this two-form is the pull-back of the symplectic
 form on the moduli space but not itself symplectic. This is due to the fact that the parametrisation of the phase space in
 terms of the variables in \eqref{gravom} exhibits a two-fold redundancy.

First, the holonomy variables $\mi,\aj,\bjj, L_\infty$ are
redundant, as they are subject to the constraint \eqref{genrell}.
The associated gauge transformations act on the holonomies by
simultaneously conjugating them  with a general $\prgr$-valued
function of
 $\mi,\aj,\bjj, L_\infty$.  Such gauge transformations arise from Chern-Simons gauge transformations of the form \eqref{astrafo} that are
 nontrivial at the basepoint $p_0$. We will see below that they also describe the transformation of the holonomies under a change of basepoint. In this section,
we will demonstrate the gauge invariance of the two-form \eqref{gravom} under conjugation
 of all holonomies. It then follows in particular
that the symplectic structure of
 the moduli space is independent of the base point. When  the gauge
is  partly fixed  by requiring the curvature at the distinguished puncture to
 lie in the Cartan subalgebra $\gothc$,
the two-form \eqref{gravom} takes a
 particularly simple form.

The second redundancy  is linked to the fact that the parametrisation
of the holonomies $\mi$ and $L_\infty$ in terms of general group
elements $g_i, g_\infty\in\prgr$, $ C\in T_\mathfrak{c}$ and
(constant)  elements $C_i\in \cup_{\iota\in
  I}T_{\gothc_\iota}$,
see \eqref{dischol2}, \eqref{boundholgamma} is not unique.
Right-multiplication of the elements $g_i$ and $g_\infty$ in
\eqref{dischol2}, \eqref{boundholgamma} by elements of the
stabiliser group of $C_i$ and $C$ yields the same expression for
the holonomies $\mi, L_\infty$. The group elements $g_i,
g_\infty\in\prgr$ are therefore only defined up to
right-multiplication with elements of the relevant stabilisers,
which gives rise to additional gauge transformations discussed at
the end of this section.

Before we enter the detailed calculations, we note that gauge
invariance manifests itself in the degeneracy of $\Omega$
restricted to the constraint surface $\mathcal C$,  i.~e.~in the
existence of a vector field $Z$ on $\mathcal{C}$  such that the
contraction of $\Omega$ with $Z$ is zero. The vector field $Z$ is
the infinitesimal generator of the gauge transformation. In our
case $\Omega$ is exact, so $\Omega =\delta \chi$ and $Z$ is an
infinitesimal gauge transformation if  and only if \bea
\label{gaugecond} \iota_Z\delta\chi =0 \Leftrightarrow {\cal
L}_Z\chi -\delta\iota_Z\chi =0, \eea where ${\cal L}_Z$ is the Lie
derivative with respect to $Z$, and we have used Cartan's formula.
It turns out that all the gauge transformations
 we encounter in this section are group actions $\rho$ of $\prgr$ or a
 subgroup of $\prgr$ on $\mathcal{C}$. For calculations we
 have found it convenient to consider finite rather than infinitesimal
 group actions. If $h(\epsilon)$ is a one-parameter subgroup of
 $\prgr$
with $h(0)=1$ and $Z$ the vector field generated by the action of
$\rho(h(\epsilon))$ on $\mathcal{C}$, then $ {\cal L}_Z\chi=0$ if
$\chi$ is invariant under pull-back with  $\rho(h(\epsilon))$;
this  turns out to be the case for most gauge transformations
considered here. The gauge invariance condition \eqref{gaugecond}
then requires that  the function
 $\iota_Z\chi =\chi(Z)$ is constant. More generally,
if $\iota_Z\chi=H$ is a non-constant function on $\mathcal {C}$, the condition
\eqref{gaugecond} is satisfied if $ {\cal L}_Z\chi =\delta H$.

The quickest way of checking if \eqref{gaugecond} holds
 is to allow the parameter $\epsilon$ in the action
$\rho(h(\epsilon))$ to be a function on the
 constraint surface, and to compute the changes in $\chi$ under the
 pull-back with $\rho(h(\epsilon))$. If $\chi$
changes according to
\bea
\label{gaugetest}
\chi \mapsto \chi +\delta(\epsilon H) +{\cal O}(\epsilon^2)
\eea
for some function $H$, it follows that $H=\chi(Z)$ and
(by specialising to a $\mathcal {C}$-independent parameter $\epsilon$)
that ${\cal L}_Z\chi =\delta H$. Hence \eqref{gaugetest} implies
\eqref{gaugecond}.
  In the following we do not specify the one-parameter subgroup
$h(\epsilon)$
but instead consider group actions of general group elements $g\in
\prgr$
which are functions on $\mathcal{ C}$. From the transformation behaviour of
$\chi$ under this action we can then read off
the behaviour under  any one-parameter subgroups with a parameter
$\epsilon$ which is a function of $\mathcal {C}$.

We start by considering the transformation of the two-form \eqref{gravom} under conjugation of all
holonomies by $g\in \prgr$ which depends on the extended phase space
variables. The transformation behaviour of the various terms in \eqref{gravom} can be stated in two
technical lemmas which can be proved using the following general formula for
conjugation in $\prgr$. Suppose $M=(u,-\mathrm{Ad}^*(u^{-1})\bj)$
and $\tilde M=(\tilde u,-\mathrm{Ad}^*(\tilde u^{-1})\tilde \pbj)$ are
related by $M=g\tilde M g^\inv$, with $g=(v,\bx)$. Then
\bea
\label{transf2}
 \bj =(1\!-\!\Ad^*(v \tilde u v^\inv))\bx\!+\!\Ad^*(v^\inv)\tilde
 \pbj,
\quad u = v \tilde u  v^\inv.
\eea
The first lemma concerns the last two terms in \eqref{intermsymp3}.
We state it in a slightly more general form.

\begin{lemma}
\label{heisenlemma}
Suppose $(\bmu,\bs)$ is an arbitrary element in the Cartan subalgebra
$\gothc$ of
 $\pralg$ (not fixed), and  the $\prgr $ elements
$g_K= (v_K,\bx_K)$ and $\tilde g_K=(\tilde v_K,\tilde \bx_K)$
are related by left-multiplication  $g_K=g\tilde
g_K$,   where  $g=(v,\bx)$ is  a function of $\tilde  g_K$ and $(\bmu,\bs)$.
 Defining
\begin{align} &K^\inv
=(u_K,-\mathrm{Ad}^*(u_K^{-1})\bj_K)=g_K(e^{-\bmu},-\bs)
g_K^{-1}\\
&\tilde K^\inv=(\tilde u_K,-\mathrm{Ad}^*(\tilde u_K^{-1})\tilde
\pbj_K)=\tilde g_K (e^{-\bmu},-\bs) \tilde g_K^{-1}
\nonumber\end{align} so that $K=g \tilde K g^{-1}$, we have the
identities \bea \label{cmhelpid0} &&\langle \bj_K,\delta v_K
v_K^\inv\rangle -\langle
\Ad^*(v_K)\bx_K,\delta \bmu\rangle \\
&=&\langle \tilde \pbj_K,\delta
\tilde v_K \tilde v_K^\inv\rangle -\langle
\Ad^*(\tilde v_K)\tilde \bx_K,\delta \bmu\rangle
+ \langle\bj_K , \delta v
v^\inv\rangle+ \langle\mathrm{Ad}^*(v)\bx, \delta \tilde u_K \tilde u_K^\inv\rangle,\nonumber\\
&=&\langle \tilde \pbj_K,\delta
\tilde v_K \tilde v_K^\inv\rangle +\langle
\Ad^*(\tilde v_K)\tilde \bx_K,\delta \bmu\rangle
 +\langle\tilde\pbj_K, v^\inv\delta v
\rangle+\langle \bx\,,\, \delta  u_K u_K^\inv\rangle.\nonumber
\eea
\end{lemma}

The lemma can be proved by straightforward calculation.
Comparison with \eqref{niceform} shows that the last two terms in
\eqref{cmhelpid0} are precisely the symplectic form on the Heisenberg
double of $\prgr$. In the terminology of Appendix \ref{heisendouble}
this lemma states that the symplectic form \eqref{favouritepot} (itself
a reduction of the Heisenberg double symplectic form)  on
the $\prgr$ element  $M$ changes by the Heisenberg double symplectic
form for $g$ and $M$ when $M$ is conjugated by a function $g$.

The second lemma concerns the
symplectic potential $\Theta$ in \eqref{gravom} and shows that it has an equal but opposite
transformation behaviour under conjugation of all the holonomies.

\begin{lemma} \label{lemma2}
If the group elements  $M_i$, $A_j$, $B_j \in\prgr$   are obtained
from $\tilde M_i,\tilde A_j,\tilde B_j$  via simultaneous
conjugation with a $\prgr$-valued function $g=(v,\bx)$ on the
extended phase space
\begin{align}
\label{ptransf}
&M_i=(u_\mi,-\mathrm{Ad}^*(u_\mi^{-1})\bj_\mi)=g\tilde\mi g^\inv=g\cdot(\tilde u_\mi,-\mathrm{Ad}^*(\tilde u_\mi^\inv)\tilde \pbj_\mi)\cdot g^\inv\\
&\aj=(u_\aj,-\mathrm{Ad}^*(u_\aj^{-1})\bj_\aj)=g\tilde\aj g^\inv=g\cdot(\tilde u_\aj,-\mathrm{Ad}^*(\tilde u_\aj^\inv)\tilde \pbj_\aj)\cdot g^\inv\nonumber\\
&\bjj=(u_\bjj,-\mathrm{Ad}^*(u_\bjj^{-1})\bj_\bjj)=g\tilde \bjj g^\inv=g\cdot(\tilde u_\bjj,-\mathrm{Ad}^*(\tilde u_\bjj^\inv )\tilde \pbj_\bjj)\cdot
g^\inv,\nonumber
\end{align}
and the variables $v_\mi$ defined in \eqref{holparam} from $\tilde
v_\mi$ via left-multiplication with the Lorentz component of this
function
\begin{align}
\label{ptransf2} v_\mi =v \tilde v_\mi
\end{align}
then
the expressions for the symplectic potential $\Theta$ in terms of these two sets of variables are related by
\begin{align}
\label{cmhelpid}
&\Theta(v_\mi,\bj_\mi, u_\aj,\bj_\aj, u_\bjj,\bj_\bjj)\\
&=\Theta(\tilde v_\mi,\tilde\pbj_\mi, \tilde u_\aj,\tilde \pbj_\aj, \tilde u_\bjj,\tilde\pbj_\bjj)-\langle\bj_{K_\ntg}\,,\, \delta v
v^\inv\rangle-\langle\mathrm{Ad}^*(v)\bx\,,\, \delta \tilde u_{K_\ntg} \tilde u_{K_\ntg}^\inv\rangle\nonumber\\
&=\Theta(\tilde v_\mi,\tilde\pbj_\mi, \tilde u_\aj,\tilde \pbj_\aj, \tilde u_\bjj,\tilde\pbj_\bjj)-\langle\tilde\pbj_{K_\ntg}\,,\, v^\inv\delta v
\rangle-\langle \bx\,,\, \delta  u_{K_\ntg} u_{K_\ntg}^\inv\rangle,\nonumber
\end{align}
where $u_{K_\ntg}$, $\bj_{K_\ntg}$, $\tilde u_{K_\ntg}$,
$\tilde\pbj_{K_\ntg}$ are given in terms of the variables
$\mi,\aj,\bjj$ and $\tilde\mi,\tilde\aj,\tilde\bjj$, respectively,
by \eqref{jtotal}, \eqref{ukidef} and \eqref{jkidef}.
\end{lemma}

Applying Lemma \ref{heisenlemma} to $K=K_\ntg$ and comparing \eqref{cmhelpid0} and \eqref{cmhelpid}, we see that the terms
arising from the transformation are equal, but with opposite sign. With the considerations above we therefore find that such transformations are gauge
symmetries of the two-form $\Omega$. Furthermore, by applying Lemma  \ref{lemma2} with
$g=g_\infty$, it is possible to transform the  holonomy variables into a new set of variables in which the two-form $\Omega$ takes a particularly simple form.
We obtain the following theorem.

\begin{theorem}
\label{resulttheor} $\quad$

\noindent 1.
Simultaneous conjugation of
the holonomies $\mi$, $i=1,\ldots, n$, $A_j,B_j$, $j=1,\ldots, g$ and
$K_{n+2g}$,  and the left-multiplication of  $(v_\mi,\bx_\mi)$,
 and $g_\infty=(v_\infty,\bx_\infty)$ by an arbitrary  element $g\in \prgr$
is a gauge transformation for the two-form \eqref{gravom}.

\noindent 2.  In terms of the variables
$\tilde v_\mi,\tilde\pbj_\mi, \tilde u_\aj,\tilde \pbj_\aj, \tilde u_\bjj,\tilde\pbj_\bjj$ defined by
\begin{align}
\label{holrel} &\tilde\mi=(\tilde u_\mi,-\Ad^*(\tilde
u_\mi^\inv)\tilde\pbj_\mi)
=g_\infty^\inv\mi g_\infty\\
&\tilde\ai=(\tilde u_\ai,-\Ad^*(\tilde
u_\ai^\inv)\tilde\pbj_\ai)=g_\infty^\inv
\ai g_\infty\nonumber\\
&\tilde\bi=(\tilde u_\bi,-\Ad^*(\tilde
u_\bi^\inv)\tilde\pbj_\bi)=g_\infty^\inv \bi g_\infty\nonumber\\
&\tilde v_\mi= v_\infty^\inv v_\mi \nonumber\end{align} the
two-form \eqref{gravom} is
\begin{align}
\label{finresult} \Omega=\tfrac{k}{2\pi}\delta\left(  \langle
(\bmu,\bs), w^\inv\delta w\rangle-\tfrac{1}{2}\langle\bmu
\delta\bs\rangle+\Theta(\tilde v_\mi,\tilde\pbj_\mi, \tilde
u_\aj,\tilde \pbj_\aj, \tilde u_\bjj,\tilde\pbj_\bjj)\right).
\end{align}

\end{theorem}

In particular, this theorem addresses the dependence of the
two-form $\Omega$ on the choice of the basepoint. Under a change
of basepoint $\pp_0\mapsto \pp_0'$,
 the holonomy of any closed curve $\gamma\in\pi_1(\surfb, \pp_0)$ with respect to basepoint $\pp_0$ transforms
 according to
\begin{align}
\label{bsptchange}
H[\gamma, \pp_0]\mapsto H[\gamma, \pp_0']=PT_{\pp_0\rightarrow \pp_0'}\cdot H[\gamma, \pp_0]\cdot PT_{\pp_0'\rightarrow \pp_0},
\end{align}
where $PT_{x\rightarrow y}$ denotes the $\prgr$-element which implements the
parallel transport from $x$ to $y$.
It follows from the first part of Theorem \eqref{resulttheor} that
expressions \eqref{gravom} for
the two-form $\Omega$ derived with respect
 to two different basepoints agree, and we have
\begin{corollary}
The two-form \eqref{gravom} does not depend on the choice of  basepoint
$\pp_0$ of the  fundamental group $\pi_1(\surf,\pp_0)$.
\end{corollary}

Note that the holonomies defined in the second part of
Theorem \ref{resulttheor} satisfy
\begin{align}
\label{cmholon}
L_\infty^\inv=[\tilde B_g,\tilde A_g^\inv]\cdots[\tilde B_1,\tilde A_1^\inv]\cdot \tilde M_n\cdots\tilde M_1=e^{\frac{2\pi}{k} D}=(e^{-\bmu}, -\bs).
\end{align}
They are therefore the holonomies in a gauge where the holonomy around the distinguished puncture is in the abelian subgroup $T_\gothc$.
This is the partial gauge fixing anticipated in the opening paragraph
of this subsection. However, as discussed above, there is a residual
gauge freedom linked to the redundancy of the parametrisation of the
holonomies $\mi$, $L_\infty$ in terms of general group elements and
elements of the subgroups $T_{\gothc_\iota}$. For the holonomy $L_\infty$, the condition \eqref{cmholon}
does not eliminate  the gauge invariance of \eqref{gravom} completely. The
condition
is preserved by transformations of the form
\begin{align}
\label{boundgt}
&w\mapsto w g\\
&\tilde X\mapsto g\tilde X g^\inv \qquad\forall\tilde X\in\{\tilde
M_1,\ldots,\tilde B_g\}\nonumber
\end{align}
for any element $g=(v,\bx)$ in the stabiliser group of
$(\bmu,\bs)$, i.e. $\Ad(v)\bmu=\bmu$ and $\Ad^*(v^\inv)\bs=\bs$.
To see that these transformation are gauge transformations of
\eqref{finresult} note the following transformation properties of
the symplectic potentials entering \eqref{finresult}:
\begin{align}\langle (\bmu,\bs), w^\inv\delta w\rangle\mapsto &\langle
 (\bmu,\bs), w^\inv\delta w\rangle+\langle\bmu,\Ad^*(v)\delta\bx \rangle+\langle\bs,v^\inv \delta
 v\rangle.
\end{align}
From Lemma \eqref{lemma2} we have
\begin{align}
\Theta(\tilde v_\mi,\tilde\pbj_\mi, \tilde u_\aj,\tilde
\pbj_\aj, \tilde u_\bjj,\tilde\pbj_\bjj)&\mapsto \\
 &\Theta(\tilde v_\mi,\tilde\pbj_\mi, \tilde u_\aj,\tilde
\pbj_\aj, \tilde u_\bjj,\tilde\pbj_\bjj)-\langle\bs,\delta v v^\inv \rangle+
 \langle\Ad^*(v)\bx ,\delta\bmu\rangle \nonumber,
\end{align}
and, using the stabiliser property of $g$, we check that \bea
\langle\Ad^*(v)\bx ,\delta\bmu\rangle = \delta
\langle\Ad^*(v)\bx,\bmu\rangle -\langle\bmu,\Ad^*(v)\delta\bx
\rangle + \langle \bs, \delta v v^{-1}\rangle -\langle \bs,
v^{-1}\delta v\rangle, \eea so that the sum $\Theta +\langle
(\bmu,\bs), w^\inv\delta w\rangle$ changes according to \bea
\Theta +\langle (\bmu,\bs), w^\inv\delta w\rangle\mapsto \Theta
+\langle (\bmu,\bs), w^\inv\delta w\rangle + \delta
\langle\Ad^*(v)\bx,\bmu\rangle. \eea For each one-parameter
subgroup $h(\epsilon)$ of the stabiliser group of $(\bmu,\bs)$
this transformation behaviour  is of the form \eqref{gaugetest},
showing that the transformations \eqref{boundgt} are indeed gauge
transformations.

Similar residual gauge transformations arise for each of the ordinary
punctures. The redundancy \eqref{carttrafoi} in the parametrisation
\eqref{tidef} of coadjoint orbits leads to the gauge transformation
\bea
\label{spingt}
\tilde v_\mi\mapsto \tilde v_\mi v
\eea
where $v$ is such that  $(v,0)$ is in the
stabiliser
group of $(\bmu_i,\bs_i)$, i.e. $\Ad(v)\bmu_i=\bmu_i$
and $\Ad^*(v^\inv)\bs_i=\bs_i$.
Under the transformation \eqref{spingt}  the symplectic
potential
\eqref{gravtheta} changes according to
\bea
\Theta(\tilde v_\mi,\tilde\pbj_\mi, \tilde u_\aj,\tilde
\pbj_\aj, \tilde u_\bjj,\tilde\pbj_\bjj)\mapsto
 \Theta(\tilde v_\mi,\tilde\pbj_\mi, \tilde u_\aj,\tilde
\pbj_\aj, \tilde u_\bjj,\tilde\pbj_\bjj)-\langle\bs_i,\delta v v^\inv \rangle
\eea
Again we see that
for each one-parameter subgroup $h(\epsilon)$ of the stabiliser group  this transformation behaviour  is of the form
\eqref{gaugetest}, showing that the transformations \eqref{spingt}
are indeed gauge transformations.

The
physical phase space $\mathcal{P}_{phys}$ is, by definition,
the quotient of the constraint
surface $\mathcal{C}$ by the gauge transformations discussed in this section.
It is difficult to
analyse $\mathcal{P}_{phys}$  in more detail without
restricting the group $\prgr$ to a smaller class or particular
example. Even for compact gauge groups it is known that the
physical phase space of Chern-Simons theory is a  manifold with
singularities, see \cite{SL}.
 Because of the
non-compactness of $\prgr$ it may happen that $\mathcal{P}_{phys}$
is  not even a Hausdorff space, see \cite{Ezawa} for  a discussion
of an example. While we study the case of the Poincar\'e group in
$(2+1)$ dimensions in more detail in \cite{we5}, here we
 only count the
dimension of $\mathcal{P}_{phys}$, assuming that it is a manifold.
The variables obtained after partial gauge fixing are the
$\prgr$-elements $w$, $\tilde\mi$ $i=1,\ldots, n$,
$\tilde\aj,\tilde\bjj $, $j=1,\ldots, g$
 and the $\gothc$-element  $(\bmu,\bs)$.
The holonomies $\tilde\mi$ are each restricted to lie in conjugacy
classes \bea \mathcal{C}_{\bmu_i,\bs_i}=\{g (e^{-\bmu_i},-\bs_i)
g^\inv |g\in \prgr\}, \eea and we have to impose the constraint
\eqref{cmholon} and divide  by the gauge transformation
\eqref{boundgt}. The variables $(\bmu, \bs)$ are good coordinates
on the physical phase space in the generic situation where  the
stabiliser group of $(\bmu,\bs)$ has the same dimension as
$\gothc$. Thus we obtain
 the dimension of  the physical phase space as follows:
\bea \text{dim}\mathcal{P}_{phys}&=&\sum_{i=1}^n\text{dim}\,
(\mathcal{C}_{\bmu_i,\bs_i}) +(2g+1)\cd\text{dim}\,
(\prgr)+\text{dim}\,
\gothc-(\text{dim}\,(\prgr)+\text{dim}\,\gothc)
\nonumber\\
&=&\sum_{i=1}^n\text{dim}\,(\mathcal{C}_{\bmu_i,\bs_i}) +2
g\cd\text{dim}\,(\prgr).  \eea

\section{Outlook and conclusion}
\label{outlook}

In this paper we introduced a new way of treating a puncture in
Chern-Simons theory and explicitly determined the symplectic
structure on the phase space of Chern-Simons theory with gauge
group $\prgr$ on $\RR \times \surfb$ when  one distinguished
puncture on $\surfb$ is treated in this way. Our main motivation
for studying gauge groups of the form $\prgr$ is the application
to (2+1)-dimensional gravity, which can be formulated as a
Chern-Simons theory
 with  gauge
groups  of the form $\prgr$ when the cosmological constant is
zero.

The application to (2+1)-dimensional gravity is described in
detail in the paper \cite{we5}, where we use a distinguished
puncture of the type considered here to model open universes. In
this context, the ordinary punctures represent particles  of mass
$\bmu_i$ and spin $\bs_i$ in a spacetime of genus $g$. The
distinguished puncture stands for the boundary of the universe at
spatial infinity, and the phase space variables $\bmu$ and $\bs$
are interpreted as total mass and spin of the universe. The
two-form \eqref{gravom} and the explicit parametrisation of the
extended phase therefore provide the basis for a detailed study of
the classical dynamics of an open universe containing an arbitrary
number of particles. Moreover, it can serve as the starting point
for an equally explicit study of (2+1)-dimensional quantum
gravity.

Although many aspects of the  model   considered here were
motivated by the application to   (2+1)-dimensional gravity, our
treatment of the distinguished puncture makes sense  for any gauge
group satisfying the technical assumptions spelled out in Sect.~2.
In a more general context, it
 may also be of interest to consider several
distinguished punctures. It is not difficult to generalise our discussion in
Sect.~\ref{combi}  to take into account several distinguished punctures
and the additional boundary components which would result from the cutting
procedure described there.

\subsection*{Acknowledgements}   BJS acknowledges support through
an EPSRC advanced fellowship during the early stages of this project.

\appendix
\section{Poisson structures associated to $\prgr$}
\label{heisenberg}

This appendix summarises properties of two symplectic structures
which play an important role in this paper. The first is the
canonical symplectic structure defined on the  cotangent bundle of
groups of the form $\prgr$, and the second is  the symplectic
structure on the Heisenberg double of $\prgr$. While the cotangent
bundle symplectic structure is defined for any Lie group, the
Heisenberg double structure is linked to the properties of $\prgr$
as a Poisson-Lie group, see  \cite{STS, AMI} and the textbook
\cite{CP}. We show how various symplectic structures which arise
in the paper can be deduced from the cotangent bundle and
Heisenberg double symplectic structures.

\subsection{Cotangent bundle of $\prgr$}
\label{cotbundle} In the parametrisation  $g=(v,\bx)\in \prgr$ the
right-invariant one-form on $\prgr$ is given by \bea \delta g
g^{-1} = (\delta v v^{-1}, \delta \bx -[\delta v v^{-1},\delta
\bx]). \eea Similarly, the left-invariant one-form is \bea g^{-1}
\delta g= (v^{-1}\delta v, \Ad^* (v)\delta \bx). \eea We  write
an element of the  Lie algebra of $\prgr$ as $T=-(\bp, \bk
)=-(p_aJ^a, k_aP^a)$ and use the canonical  inner product
$\langle, \rangle$ on $\pralg$ to identify the dual  $\pralg$ with
the Lie algebra $\pralg$. Then the canonical symplectic structure
on the cotangent bundle of $\prgr$ is \bea \label{universal}
\Omega_0 =\frac 1 2 (\langle \delta T,\delta g g^{-1} \rangle+
\langle \delta \tilde T, g^{-1} \delta g\rangle ), \eea where
$\tilde T =g^{-1} Tg$. This expression is useful for understanding
the relationship with the Heisenberg double. For calculations we
use \bea \Omega_0=\delta \langle  T ,\delta g g^{-1} \rangle =
\delta \langle
 \tilde T, g^{-1} \delta g\rangle
\eea
to find
\bea
\label{cotbundlesymp}
\Omega_0=-\delta \left(\langle\bp, \delta \bx\rangle +\langle
  \bk-[\bx,\bp], \delta v v^{-1}\rangle\right).
\eea

Two further symplectic structures can be obtained from the cotangent
bundle by reduction. Pick a  Cartan subalgebra of $\prgr$, denote
it  by $\gothc$ and
and write  $(\bmu,\bs)$ for an element in $\gothc$. We obtain a
symplectic structure on $\prgr \times \gothc$ from  \eqref{universal}
by restricting $\tilde T$ to lie in $\gothc$. Parametrising
\bea
\label{ttpara}
 \tilde T = -(\bmu,\bs)
\eea
so that
\bea
\label{tpara}
T=-g(\bmu,\bs)g^\inv
\eea
we now have the formulae
\begin{align}
\label{pformula} &\bp = \Ad(v) \bmu \\
\label{kformula} &\bk
=[\bx,\bp] +\Ad^*(v^{-1})\bs \end{align}
 for
$\bp$ and $\bk$ in terms  of  $g=(v,\bx)$ and $(\bmu,\bs)$. Inserting
\eqref{ttpara} and \eqref{tpara}
 into \eqref{universal} one obtains
\bea
\omega_0=-\delta \langle(\bmu,\bs) , g^{-1}\delta g\rangle =
\delta \langle T, \delta g g^{-1}\rangle.
\eea
In terms of \eqref{pformula} and \eqref{kformula}  this can be
written as
\bea
\label{omegazero}
\omega_0&=&
-\delta (\langle
  \bs,  v^{-1}\delta  v\rangle
 + \langle \bp,\delta \bx \rangle) \\
&=&-\delta (\langle
  \bk, \delta v v^{-1}\rangle - \langle\Ad(v)\delta\bmu ,\bx\rangle).\nonumber
\eea

If we  require $(\bmu,\bs)\in\gothc$ to be constant, the form
\eqref{omegazero} becomes degenerate and ceases to be symplectic.
It is, however, the pull-back of a symplectic form, namely the
canonical symplectic form  on the coadjoint orbit ${\cal
O}_{\bmu,\bs}=\{ g  (\bmu,\bs)g^\inv|\, g\in \prgr\}$. These
orbits are the symplectic leaves of the canonical Poisson
structure on the dual of the Lie algebra of $\prgr$.

\subsection{Heisenberg double of $\prgr$}
\label{heisendouble}
The Heisenberg double of a Poisson-Lie group is a group with a Poisson
structure, but it is not a Poisson-Lie group \cite{STS,AMI}.
The Heisenberg double  $D_+(\prgr)$ of the  semi-direct product group $\prgr$
can be identified with the Cartesian product
 $\left(\prgr\right) \times \left(\prgr\right)$ as a group. In order to
describe its Poisson structure we need the dual Poisson-Lie group
of $\prgr$, which is isomorphic to the direct product
$G\times\gothg^*$ \cite{we2}. According to the general theory of
Poisson-Lie groups \cite{CP} there exist group homomorphisms
 $S,S_\sigma:G\times\gothg^* \rightarrow \prgr$
such that
\bea
\label{diffeo}
Z: G\times\gothg^* &\rightarrow& \prgr \\
h^*&\mapsto& S(h^*)(S_\sigma (h^*))^{-1} \nonumber
\eea
is a diffeomorphism, at least locally. In our case, we parametrise
elements
$h^*\in G\times \gothg^*$ as $h^*=(u,-\bj)$ and find
\bea
S(u,-\bj)=(u,0)\quad S_\sigma (u,-\bj )=(1,\bj).
\eea
Hence
\bea
\label{diffeoo}
Z: (u,-\bj)\mapsto  S(u,-\bj)(S_\sigma (u,-\bj ))^{-1}=(u,-\Ad^{*}
 (u^\inv)\bj),
\eea
which is a global diffeomorphism.

Now turning to the Heisenberg double, we have homomorphisms
\bea
\prgr  &\rightarrow&  D_+(\prgr) \\
g&\mapsto& (g,g) \nonumber
\eea
and
\bea
G\times \gothg^* &\rightarrow&  D_+(\prgr) \\
h^*=(u,-\bj) &\mapsto& (S(h^*),S_\sigma(h^*))= ((u,0),(1,\bj)).
\nonumber \eea The Heisenberg double of $\prgr$ is factorisable,
so we can define  further elements   $h=(w,\by)\in \prgr $ and
$g^*=(\tilde u,- \tilde \pbj) \in G \times \gothg^*$ via the
relation \bea \label{factor} g g^*=h^* h \eea in $D_+(\prgr)$. In
particular this implies that $h^*$ is related to $g^*$ via a
dressing  transformation, which is  the $\prgr$-conjugation action
on the  $\prgr$-element associated via $Z$: \bea
g(Z(g^*))g^{-1}=Z(h^*) \eea or, in components: \bea \label{orbit}
u=v\tilde uv^{-1}\quad \bj=(1-\Ad^*(u))\bx +\Ad^*( v^{-1})\tilde
\pbj. \eea For the $\prgr$ elements $h$ and $g$,  the relation
\eqref{factor}
 implies
\bea \label{simp} w=v\qquad \by =\Ad^* (u) \bx. \eea Following the
notation used by Alekseev and Malkin in \cite{AMI} , the
symplectic structure on the Heisenberg double is \bea \Omega_H &=&
\frac 1 2 \left(\langle \delta h^* (h^*)^{-1} , \delta g
  g^{-1}
\rangle + \langle (g^*)^{-1}\delta g^*, h^{-1} \delta h \rangle
\right), \eea and with our parametrisation of the elements
$h,h^*,g,g^*$ we have \bea \Omega_H &=&  \frac 1 2( \langle \delta
u u^{-1},-\delta \bj),\delta
  g g^{-1}\rangle  + \langle (\tilde u^{-1}\delta \tilde u,
-\delta \tilde \pbj), h^{-1}\delta
  h \rangle ) \\
 &= &\frac 1 2 (\langle \delta u u^{-1},\delta \bx -
[\delta v v^{-1},\bx]\rangle  -  \langle \delta \bj, \delta v
v^{-1}\rangle \nonumber \\
&& + \langle  \tilde u^{-1}\delta \tilde u,\Ad^* (w)
\delta \by\rangle - \langle
\delta \tilde \pbj, w^{-1} \delta w\rangle). \nonumber
\eea
Using \eqref{orbit} and \eqref{simp} to write
$g^*$ and $h$ in terms of $g$ and $h^*$
 we find, after some
algebra:
\bea
\label{niceform}
\Omega_H & =& -\delta \left( \langle \bj - (1-\Ad^* (u)) \bx ,  \delta v
v^{-1}\rangle + \langle \bx, \delta u u^{-1}\rangle \right) \\
&=& -\delta \left( \langle \bj, \delta v
v^{-1}\rangle  + \langle \Ad^*(v) \bx, \delta \tilde u \tilde
u^{-1}\rangle \right).\nonumber
\eea
Note that we  recover the symplectic  structure
\eqref{cotbundlesymp}  of the
 cotangent bundle $T^*(\prgr)$  if  we
expand $u=e^{-p_aJ^a}$ and keep linear terms in $\bp$.

In analogy with the cotangent bundle symplectic structure we briefly
describe two symplectic structures derived from the Heisenberg double
structure. The first is defined on the product of $\prgr$ with the
Cartan subalgebra $\gothc$, assumed to be abelian.
Restricting $h^*$ to be of the form
 $(\tilde u, -\tilde \pbj)=(e^{-\bmu}, -\bs )$ with $(\bmu,\bs)\in
 \gothc$ arbitrary yields
\bea
\label{favouritepot}
\omega_H &=& -\delta (\langle \bs,  v^{-1}\delta v\rangle +\langle
\bx,\delta u  u^{-1}\rangle )
 \\
&=&  -\delta (\langle \bj, \delta v v^{-1}\rangle -\langle\bx,
 \Ad(v) \delta\bmu\rangle)\nonumber,
\eea
where now
\bea
\label{ujform}
\bj = (1-\Ad^*( u))\bx +  \Ad^*( v^{-1}) \bs\quad   \text{and} \quad
u=ve^{-\bmu}v^{-1}.
\eea

Finally, we obtain a degenerate two-form by setting
 $\bmu$ and $\bs$  to  constant values. This two-form is the
 pull-back of a canonical symplectic structure on the conjugacy
 classes of $\prgr$ \cite{we2}.
According to the general
 theory of Poisson-Lie groups \cite{CP} these conjugacy classes
  are the symplectic
leaves of  the dual Poisson-Lie group $G\times \gothg^*$. They are
analogous to the coadjoint orbits discussed at the end of the previous
subsection.

\section{Alekseev and Malkin's description of the symplectic structure}
\label{alekseevmalkin} In \cite{AMII}, Alekseev and Malkin  study
the moduli space of flat $H$-connections for simple complex  Lie
groups $H$ (or their compact real  forms)  on a two-dimensional
surface $\surf$ of genus $g$ with $n$ punctures. Most of their
results apply also to the case $H=\prgr$ studied in the present
paper. In this appendix we summarise some of the results from
\cite{AMII} required in our calculations, using our notation, and
indicate modifications where necessary.
 The first  applies without
 change
to our situation.

\begin{lemma}(Lemma 2 \cite{AMII})

\label{gflemma}
Let $P\subset\surf$ be region of $\surf$ on which the $H$-connection $A_S$
takes the form
\begin{align}
\label{gfexpr} A_S=\gamma B \gamma^\inv +\gamma d\gamma^\inv
\end{align}
where $B$ is a gauge field on $P$ with curvature $F_B=dB+B\wedge B$
and
$\gamma:\surf\rightarrow H$. Then, the canonical symplectic form on $P$ is
given by
\begin{align}
\label{sympformexp} \int_P \langle\delta A_S\wedge\delta
A_S\rangle=\int_P
\langle \delta B\wedge\delta B\rangle+2\delta\langle
F_B\,,\,\delta\gamma^\inv \gamma\rangle+\int_{\partial P} \langle
\delta
\gamma^\inv \gamma\,,\, d(\delta\gamma^\inv\gamma)\rangle-2\delta\langle
B\,,\,\delta\gamma^\inv\gamma \rangle.
\end{align}
\end{lemma}

The other results we require from \cite{AMII} concern the
evaluation of the contribution of the ordinary punctures and
handles to the  two-form \eqref{symp}. As explained in
Sect.~\ref{combi}, these contributions can be expressed as
boundary integrals along the curves $m_i, a_j,b_j$, which can be
evaluated by means of a continuity condition and related to the
corresponding holonomies $M_i,A_j,B_j$. We summarise the results
obtained in \cite{AMII} and sketch their derivation.

We start by considering the ordinary punctures. Expression
\eqref{symp} for the two-form $\Omega$ involves two integrals
along the  curve $m_i$. The sum of these, in the following denoted
by $\Omega_\mi$ is the contribution of the $i$-the puncture to the
symplectic structure.  Again the result applies to $H=\prgr$
without change and is given by the following lemma

\begin{lemma} (Lemma 3, \cite{AMII})
\label{amlemma3}

Let
\bea
\label{mborde}
\Omega_\mi=\int_{m_i} \langle\delta \gamma_0^\inv \gamma_0 d\left(
\delta \gamma_0^\inv\gamma_0\right)\rangle-\int_{m_i} \langle\delta
\gamma_\mi^\inv \gamma_\mi d\left( \delta \gamma_\mi^\inv \gamma_\mi
\right)\rangle-\tfrac{2}{k}\delta \langle D_i\delta \gamma_\mi^\inv
\gamma_\mi\rangle d\phi_i
\eea
with $\gamma_0,\gamma_\mi$ related by the overlap condition
\bea
\label{mover2}
\gamma_0^\inv|_{m_i}=N_\mi D_\mi(\phi_i)\gamma_\mi^\inv|_{m_i},
\eea
 where $C_\mi(\phi_i)=\exp(\tfrac{1}{k}D_i\phi_i)$
 and $N_\mi$ is an arbitrary but  constant element of the gauge group
$\prgr$. Denote starting and endpoint of $m_i$ by $p_{i-1}$ and $p_i$ and define
\begin{align}
\label{kgcdef}
&K_i=\gamma_0^\inv(p_i)\quad K_{i-1}=\gamma_0^\inv(p_{i-1}) & g_i=\gamma_\mi(p_i)=\gamma_\mi(p_{i-1}).
\end{align}
Then, the holonomy $M_i$ along $m_i$ is given by
\begin{align}
\label{miapp} M_i=g_i C_i g_i^\inv=K_i^\inv K_{i-1},
\end{align}
and the two-form \eqref{mborde}  can be expressed as
\bea \label{mcontrib}
\Omega_\mi=\frac{k}{4\pi}\langle C_i g_i^\inv\delta g_i C_i^\inv\wedge
g_i^\inv\delta g_i\rangle -\langle \delta K_i K_i^\inv\wedge\delta K_{i-1}
K_{i-1}^\inv
\rangle.
\eea
\end{lemma}

{\bf Proof \cite{AMII}:}

The overlap condition
\eqref{mover2} implies
\begin{align}
\label{moverl2}
\langle\delta\gamma_0^\inv\gamma_0
d\left(\delta\gamma_0^\inv\gamma_0\right)
\rangle|_{m_i}=&d\langle\delta N_\mi
N_\mi^\inv\;\delta\gamma_0^\inv\gamma_0
\rangle|_{m_i}+\langle\delta\gamma_\mi^\inv\gamma_\mi d
\left(\delta\gamma_\mi^\inv\gamma_\mi\right)\rangle|_{m_i}\\
&-\tfrac{2}{k}\delta \langle D_i d\phi_i\;\delta\gamma_\mi^\inv
\gamma_\mi\rangle|_{m_i}.\nonumber
\end{align}
Using this identity in \eqref{mborde}, one obtains
\bea
\label{mborder}
\Omega_\mi=\left[\langle\delta N_\mi N_\mi^\inv\wedge
  \delta\gamma_0^\inv\gamma_0
\rangle\right]_{\pp_{i-1}}^{\pp_i}=-\left[\langle \gamma_\mi\delta
\gamma_\mi^\inv\wedge \gamma_0 \delta \gamma_0^\inv\rangle\right]_{\pp_{i-1}}^{\pp_i},
\eea
where the last equality follows by using the overlap condition
\eqref{moverl2}  to evaluate the term $\delta N_\mi N_\mi^\inv$.
Denoting the values of $\gamma_0$ and $\gamma_\mi$ at the endpoints by, respectively, $K_i$, $K_{i-1}$ and $g_i$ as in \eqref{kgcdef}
 and applying \eqref{miapp} repeatedly one obtains expression \eqref{mcontrib}.\hfill$\Box$

Next we turn to the contribution of  the handles. For each handle,
there are two non-contractible loops $a_j$ and $b_j$
whose homotopy classes are generators of the fundamental group
 $\pi_1(\surfb,\pp_0)$. When cutting $\surfb$ to produce the Polygon
$P_0$ each of these curves appears twice in the boundary of $P_0$,
with opposite orientation. As in the main text after \eqref{hoverl}
we denote a generic generator $\in\{a_1,b_1,\ldots,a_g,b_g\}$
by $x$ and use  the notation $\gamma_0'$ and
$\gamma_0''$
for the restriction of  trivialising gauge transformation $\gamma_0$
to the two boundary
components of $P_0$ corresponding to $x$. $\gamma_0'$ and $\gamma_0''$ then satisfy the continuity condition
\bea
\label{hover2}
\left(\gamma'_0\right)^{-1}|_x=N_x \lbr \gamma''_0 \rbr^{-1}|_x\quad\text{with}\quad dN_x=0,
\eea
and, denoting  the
endpoints  of $x$ by
$y_1,y_2$,
 the contribution to the two-form
 \eqref{symp} is given by
\bea
\label{hborde}
\Omega_X=\int_{y_1}^{y_2} \langle\delta \lbr\gamma'_0\rbr^{-1}
\gamma'_0
d\left( \delta
  \lbr\gamma'_0\rbr^{-1}\gamma'_0\right)\rangle-\int_{y_1}^{y_2}
\langle\delta \lbr\gamma''_0\rbr^{-1} \gamma''_0 d\left( \delta
  \lbr\gamma''_0\rbr^{-1}
\gamma''_0\right)\rangle. \eea The full contribution
$\Omega_{H_i}$ of a handle is obtained by adding \eqref{hborde}
for the corresponding $a$- and $b$-cycle and can be evaluated as
follows.

\begin{lemma} (Lemma 4, \cite{AMII})

\label{amlemma4}
Let  $\Omega_\ai$ and $\Omega_\bi$ be defined according to \eqref{hborde} with endpoints as given in Fig. \ref{cut2} and set
\begin{align}
\label{hborde2}
\Omega_{H_i}=\Omega_\ai+\Omega_\bi.
\end{align}
Let the values of the function $\gamma_0$ at the endpoints $p_i$
be given  as in \eqref{cornerval2}
\begin{align}
\label{cornerval22}
&\gamma_0(\pp_{n+4j-3})=K_{n+2j-1}^\inv:=A_j[B_{j-1},A_{j-1}^\inv]\cdots[B_1,A_1^\inv]M_n\cdots M_1\\
&\gamma_0(\pp_{n+4j-2})=B_j^\inv A_j[B_{j-1},A_{j-1}^\inv]\cdots[B_1,A_1^\inv]M_n\cdots M_1\nonumber\\
&\gamma_0(\pp_{n+4j-1})=A_j^\inv B_j^\inv A_j[B_{j-1},A_{j-1}^\inv]\cdots[B_1,A_1^\inv]M_n\cdots M_1\nonumber\\
&\gamma_0(\pp_{n+4j})=K_{n+2j}^\inv:=[B_j,\aj^\inv]\cdots[B_1,A_1^\inv]M_n\cdots
M_1\quad\text{for}\; 1\leq j\leq g\nonumber,
\end{align}
and parametrise the holonomies $A_i,B_i$, $i=1,\ldots, g$
 in terms of variables
$g_{n+2i-1}, g_{n+2i}\in H$
 and $C_{n+i} $ in  a fixed Cartan subgroup via
 \begin{align}
\label{handhol02app}
&A_i=\gamma_0(\pp_{n+4i-3})\gamma_0^\inv(\pp_{n+4(i-1)})=\gamma_0(\pp_{n+4i-2})\gamma_0^\inv(\pp_{n+4i-1})=g_{n+2i-1}C_{n+i}^\inv g_{n+2i-1}^\inv\\
&B_i=\gamma_0(\pp_{n+4i})\gamma^\inv_0(\pp_{n+4i-1})=\gamma_0(\pp_{n+4i-3})\gamma_0^\inv(\pp_{n+4i-2})=g_{n+2i}g_{n+2i-1}^\inv.\nonumber
\end{align}
Then, the two-form \eqref{hborde2} is given  by
\begin{align}
\label{hcontribcsa}
\Omega_{H_i}=&\Omega_\ai+\Omega_\bi\\
=&\frac{k}{4\pi}\langle C_{n+i} g_{n+2i-1}^\inv\delta g_{n+2i-1}
C_{n+i}^\inv
\wedge g_{n+2i-1}^\inv\delta g_{n+2i-1}\rangle\nonumber\\
+&\frac{k}{4\pi}\langle C_{n+i}^\inv g_{n+2i}^\inv\delta g_{n+2i}
C_{n+i}\wedge
g_{n+2i}^\inv\delta g_{n+2i}\rangle\nonumber\\
-&\frac{k}{2\pi}\langle \delta
C_{n+i}C_{n+i}^\inv\wedge\left(g_{n+2i-1}^\inv
\delta g_{n+2i-1}-g_{n+2i}^\inv\delta g_{n+2i}\right)\rangle\nonumber\\
-&\frac{k}{4\pi}\left(\langle \delta K_{n+2i-1} K_{n+2i-1}^\inv\wedge\delta
K_{n+2(i-1)}
K_{n+2(i-1)}^\inv\rangle +\langle\delta K_{n+2i}
K_{n+2i}^\inv\wedge\delta K_{n+2i-1} K_{n+2i-1}^\inv\rangle\right).\nonumber
\end{align}

\end{lemma}

{\bf Proof \cite{AMII}:}

The overlap condition \eqref{hover2} implies \bea
\langle\delta\lbr\gamma'_0\rbr^{-1}\gamma'_0 d
(\delta\lbr\gamma'_0\rbr^{-1} \gamma'_0 )\rangle|_x= d\langle
\delta N_x N_x^\inv ,\delta\lbr\gamma'_0\rbr^{-1}
\gamma'_0\rangle|_x + \langle\delta\lbr\gamma''_0\rbr^{-1}
\gamma''_0, d(\delta\lbr\gamma''_0\rbr^{-1} \gamma''_0)\rangle|_x,
\eea and inserting this expression into \eqref{hborde} gives \bea
\label{hborder} \Omega_X= \left[\langle\delta N_xN_x^\inv\wedge
  \delta\lbr \gamma'_0\rbr^{-1}\gamma'_0
\rangle\right]_{y_1}^{y_2}=-\left[\langle \gamma''_0\delta
\lbr\gamma''_0\rbr^{-1}\wedge \gamma'_0  \delta
\lbr\gamma'_0\rbr^{-1} \rangle\right]_{y_1}^{y_2}. \eea After a
rather lengthy calculation which repeatedly makes use of
\eqref{handhol02app}, \eqref{cornerval2} and adding the
contributions of the
 generators $a_i$ and $b_i$ one obtains expression \eqref{hcontrib}.\hfill$\Box$

As explained in Sect.~2,
in  groups of the form  $\prgr$
we cannot assume without loss of generality
that the holomomy variables
 $A_i$ and $B_i$,
$i=1,\ldots, g$ can be parametrised in terms of elements of a
given Cartan subgroup, or even elements of a finite union  of
Cartan subgroups, via \eqref{handhol02app}. We have therefore
repeated the calculations without making use of such a
parametrisation. The result is

\begin{lemma}
\label{ourcase} With  notation in Lemma \ref{amlemma4}, the the
two-form \eqref{hborde2} can be written as
\begin{align}
\label{hcontrib}
\Omega_{H_i}=&\Omega_\ai+\Omega_\bi\\
=-&\frac{k}{4\pi}\sum_{i=1}^{g} \left(\langle A_i^\inv \delta
A_i\wedge B_i^\inv \delta B_i\rangle + \langle
\delta(B_iA_iB_i^\inv)B_iA_i^\inv B_i^\inv \wedge \delta B_i\,
B_i^\inv \rangle\right) \nonumber\\ -&\frac{k}{4\pi}\left(\langle
\delta K_{n+2i-1} K_{n+2i-1}^\inv\wedge\delta K_{n+2i-2}
K_{n+2i-2}^\inv\rangle +\langle\delta K_{n+2i}
K_{n+2i}^\inv\wedge\delta K_{n+2i-1}
K_{n+2i-1}^\inv\rangle\right).\nonumber
\end{align}
\end{lemma}

This is the formula  used  in the main text of the paper.

\end{document}